\documentclass[a4paper, 12pt]{article}
\usepackage[utf8]{inputenc}
\usepackage[a4paper, width=150mm, top=25mm, bottom=25mm]{geometry}
\usepackage{mathptmx}
\usepackage{amsmath,amssymb}
\usepackage[mathscr]{euscript}
\usepackage{graphicx}
\usepackage{booktabs}
\usepackage{makecell}
\usepackage{setspace}
\usepackage{helvet}
\usepackage{hyperref}
\usepackage{nameref}
\usepackage{caption,subcaption}
\usepackage{float}
\usepackage{svg}
\usepackage{booktabs}
\usepackage{diagbox}
\usepackage{threeparttable}
\usepackage{rotating}
\usepackage{pdflscape}
\usepackage{geometry}
\usepackage{multirow}
\usepackage{tikz, tikz-cd}
\usetikzlibrary{shapes.geometric, arrows, arrows.meta, bending}
\usepackage{algorithm}
\usepackage[noend]{algpseudocode}
\usepackage{natbib}
\usepackage{fontawesome}
\usepackage{dsfont}

\DeclareMathOperator*{\argmin}{arg\,min}
\DeclareMathOperator*{\argmax}{arg\,max}

\makeatletter
\renewcommand{\maketitle}{\bgroup\setlength{\parindent}{0pt}
\begin{flushleft}
  \textbf{\@title}
  \newline
  
  \@author
\end{flushleft}\egroup
}
\makeatother

\title{\LARGE Model Calibration and Validation From A Statistical Inference Perspective}
\author{
Samson Ting$^{a,*}$, Thomas Lymburn$^{a, b}$, Thomas Stemler$^{a}$, Yuchao Sun$^{a, b}$,  Michael Small$^{a, c}$\\
$^{a}$The Complex Systems Group, Department of Mathematics and Statistics, The University of Western Australia, Perth, Western Australia, Australia\\
$^{b}$Planning and Transport Research Centre (PATREC), The University of Western Australia, Perth, Western Australia, Australia\\
$^{c}$Mineral Resources, Commonwealth Scientific and Industrial Research Organisation (CSIRO), Kensington, Western Australia\\
*Corresponding author. Email address: samson.ting@research.uwa.edu.au
}

\begin{document}
\maketitle

\section*{Abstract} \label{abstract}
Despite the general consensus in transport research community that model calibration and validation are necessary to enhance model predictive performance, there exist significant inconsistencies in the literature. This is primarily due to a lack of consistent definitions, and a unified and statistically sound framework. In this paper, we provide a general and rigorous formulation of the model calibration and validation problem, and highlight its relation to statistical inference. We also conduct a comprehensive review of the steps and challenges involved, as well as point out inconsistencies, before providing suggestions on improving the current practices. This paper is intended to help the practitioners better understand the nature of model calibration and validation, and to promote statistically rigorous and correct practices. Although the examples are drawn from a transport research background --- and that is our target audience --- the content in this paper is equally applicable to other modelling contexts.\\
\newline
\noindent
Keywords: \emph{Parameters Estimation, Optimisation, Bayesian Inference, Bias-Variance Tradeoff, Uncertainty Quantification}


\section{Introduction} \label{introduction}
Model calibration and validation are widely recognised as necessary steps before the model can be trusted to make predictions. The most commonly adopted definitions for model calibration and model validation respectively are along the lines of \emph{the process of searching for an optimal set of parameters values that minimises discrepancy between model output and observed output} \citep{hourdakis2003practical, park2003microscopic, ma2007calibration} and \emph{the process of ensuring the model approximates reality sufficiently well} \citep{benekohal1991procedure, wu2003validation, toledo2004statistical} respectively. 

Although these definitions are intuitive, the inconsistent and sometimes contradictory practices found in existing transport modelling literature suggest the need for a more rigorous and unifying perspective on model calibration and validation. For instance, some studies such as \citep{toledo2004calibration, gagnon2008calibration} assessed model fit with calibration error but this could easily result in over-fitting. Although other studies recognise that validation error is the more appropriate metric, we also found significant differences in their interpretations of validation and consequently their implementations (see section~\ref{model_validation}).

In the authors' opinions, these intuitive one line definitions grossly oversimplify the tasks and hide a lot of subtleties, which if not clarified can lead to contradictory practices. This is further exacerbated by even less consistent definitions of terminologies. The terms \emph{model, validation} and \emph{parameters} are mentioned in almost every research concerning model calibration and validation yet rarely were they defined concisely. While this might seem pedantic, some crucial questions facing modellers are: is the model in transport research referring to just the traffic simulator or does it further include the observation noise model that describes the data collection process? Does model validation necessarily involve comparing model performance on a different data set to that used in calibration? What is the difference between inputs and parameters when both are simply arguments to a model, and is the distinction necessarily due to observability?

This research is motivated to answer these questions and many more. This paper addresses the research gap by reviewing the existing practices for model calibration and validation, and more importantly providing a unifying and statistically coherent perspective. The aim is to provide a sufficiently general road map for model calibration and validation, and to promote statistically rigorous and correct practices. Our focus is on how model calibration and validation can help to improve model predictive performance and/or explainability regardless of how such knowledge might be used to inform subsequent decision-making.

The contribution of this paper is two-fold. Firstly, we provide a thorough review of the current practices on model calibration and validation in transport modelling, and identify the sources of inconsistencies, malpractices and limitations. Secondly, this then allows us to establish a rigorous formulation of model calibration and validation that is sufficiently general, using concise terminologies from statistical inference. The central ideas are drawn from statistical learning, statistical inference, estimation theory and machine learning but we also explicitly point out the subtle differences that might prohibit ideas from different fields to be fully transferable. Although our examples are from a transport research context, the content of this paper is equally applicable to other modelling contexts. 

The paper is structured as follows: Section~\ref{problem_formulation} introduces motivating examples from traffic modelling and defines all relevant terms and notations. Section~\ref{model_calibration} formulates the model calibration problem in terms of estimators, which extends beyond the conventional optimisation-based approach. Section~\ref{model_validation} follows up with model validation and highlights the inconsistencies in existing practices. Section~\ref{bayesian_inference} discusses how statistical inference and more precisely Bayesian inference complements model calibration and validation. Section~\ref{improvements} provides some suggestions on the research gaps that can improve the current practices. Section~\ref{conclusions} concludes the paper.


\section{Examples and Definitions} \label{problem_formulation}

This section introduces some motivational examples from traffic modelling and defines all relevant terms and notations (see \nameref{appendix_a} for summary) used throughout the paper.

    \subsection{Traffic Modelling Examples}
    The following three modelling problems common in transport research were selected as illustrating examples based on diversity in scale and complexity.

        \subsubsection*{Example 1: Car Following Model}
\begin{figure}[htbp!]
            \centering
            \begin{tikzpicture}[scale=0.6]

    \draw[very thick, rounded corners=0.5ex,fill=black!20!blue!20!white,thick]  (2.5,1.8) -- ++(0.7,0.7) -- ++(1.6,0) -- ++(1,-0.7) -- (2.5,1.8);
    \draw[thick]  (4.2,1.8) -- (4.2,2.5);
    \draw[fill=blue!80!white, rounded corners=1.2ex,very thick] (1.5, .5) -- ++ (0, 1.3) -- ++(4.3,0) -- ++(1, -0.3) -- ++(0, -1) -- (1.5, .5) -- cycle;
    \draw[draw=black,fill=gray!50,thick] (2.75,.5) circle (.5);
    \draw[draw=black,fill=gray!50,thick] (5.5,.5) circle (.5);
    \draw (4,-1.5) node [anchor=south]{$x_n(t), v_n(t), a_n(t)$};
    \draw[draw=black,fill=gray!80,semithick] (2.75,.5) circle (.4);
    \draw[draw=black,fill=gray!80,semithick] (5.5,.5) circle (.4);

    \draw[very thick, rounded corners=0.5ex,fill=black!20!blue!20!white,thick]  (12.5,1.8) -- ++(0.7,0.7) -- ++(1.6,0) -- ++(1,-0.7) -- (12.5,1.8);
    \draw[thick]  (14.2,1.8) -- (14.2,2.5);
    \draw[fill=red!80!white, rounded corners=1.2ex,very thick] (11.5, .5) -- ++ (0, 1.3) -- ++(4.3,0) -- ++(1, -0.3) -- ++(0, -1) -- (11.5, .5) -- cycle;
    \draw[draw=black,fill=gray!50,thick] (12.75,.5) circle (.5);
    \draw[draw=black,fill=gray!50,thick] (15.5,.5) circle (.5);
    \draw (14,-1.5) node [anchor=south]{$x_{n-1}(t), v_{n-1}(t), a_{n-1}(t)$};
    \draw[draw=black,fill=gray!80,semithick] (12.75,.5) circle (.4);
    \draw[draw=black,fill=gray!80,semithick] (15.5,.5) circle (.4);

    \draw[-,semithick] (-.5,0) -- (24,0);
    \draw (18.5,0) node {};

    \draw[semithick] (6.8, 3.5) -- ++ (0, -0.5);
    \draw[semithick] (6.8, 3.25) -- ++ (10, 0) node[midway, above,align=center] {$\Delta x_n(t) = x_{n-1}(t) - x_{n}(t)$};
    \draw[semithick] (16.8, 3.5) -- ++ (0, -0.5);

    \draw[black, -{Triangle[width=18pt,length=12pt]}, line width=8pt](20,1.35) -- ++ (2, 0);

\end{tikzpicture}
            \caption{Illustration of car following model. The red (blue) vehicle is the leader (follower). The subscript means the follower is the $n^{\text{th}}$ vehicle in a platoon and the car following model is applied between every pair of leader and follower.}
            \label{fig:car_following}
        \end{figure}
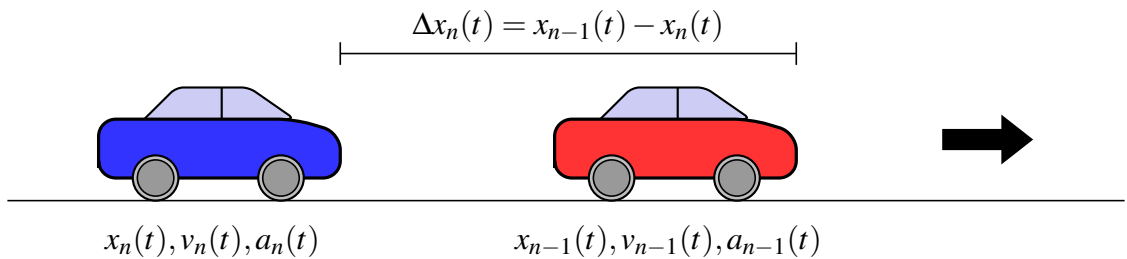

        The car following model describes the longitudinal movement of a pair of vehicles, see Figure~\ref{fig:car_following}. The vehicles' state variables are the usual kinematic variables, i.e. displacement $s(t)$, velocity $v(t)$ and acceleration $a(t)$. Given the current states of the leader and follower, the aim is to predict the follower's response at the next time step $t + \Delta t$. The parameters of interest are usually driver behavioural parameters as is the case for the popular Gipps' CF model \citep{gipps1981behavioural}. Although car following only models local behaviour, it is an important building block of the more powerful network simulation model and hence has been studied extensively.

        \subsubsection*{Example 2: Roundabout Capacity Model}
\begin{figure}[htbp!]
            \begin{subfigure}{0.5\textwidth}
            \centering
              \begin{tikzpicture}[scale=0.8]
    \coordinate (centre) at (0,0);
    \draw (centre) circle (0.75);
    \draw[dashed] (centre) circle (1.1);
    
    \draw (30:1.45) arc (30:60:1.45) arc (240:180:0.2) -- ++(0,1.5);
    \draw (30:1.45) arc (200:270:0.2) -- ++(1.5,0);
    \draw (150:1.45) arc (150:120:1.45) arc (300:360:0.2) -- ++(0,1.5);
    \draw (150:1.45) arc (330:270:0.2) -- ++(-1.5,0);
    \draw (-30:1.45) arc (150:90:0.2) -- ++(1.5,0);
    \draw (-30:1.45) arc (-30:-60:1.45) arc (-240:-180:0.2) -- ++(0,-1.5);
    \draw (210:1.45) arc (30:90:0.2) -- ++(-1.5,0);
    \draw (210:1.45) arc (210:240:1.45) arc (60:0:0.2) -- ++(0,-1.5);
    
    \draw [densely dotted] (65:1.45) arc (65:90:1.45);
    \draw [double] (90:1.45) -- ++(0,1.5);
    \draw (77:1.45) -- ++(0,0.5) coordinate (A) edge[dashed] ++(0,1);
    \draw [dashed] (103:1.45) -- ++(0,1.5);
    \draw [dashed] (103:1.45) arc (-10:-40:1);
    
    \draw [densely dotted] (-25:1.45) arc (-25:0:1.45);
    \draw [double] (0:1.45) -- ++(1.5,0);
    \draw (-13:1.45) -- ++(0.5,0) coordinate (B) edge[dashed] ++(1,0);
    \draw [dashed] (13:1.45) -- ++(1.5,0);
    \draw [dashed] (13:1.45) arc (260:230:1);
    
    \draw [densely dotted] (245:1.45) arc (245:270:1.45);
    \draw [double] (270:1.45) -- ++(0,-1.5);
    \draw (257:1.45) -- ++(0,-0.5) coordinate (C) edge[dashed] ++(0,-1);
    \draw [dashed] (283:1.45) -- ++(0,-1.5);
    \draw [dashed] (283:1.45) arc (170:140:1);
    
    \draw [densely dotted] (155:1.45) arc (155:180:1.45);
    \draw [double] (180:1.45) -- ++(-1.5,0);
    \draw (167:1.45) -- ++(-0.5,0) coordinate (D) edge[dashed] ++(-1,0);
    \draw [dashed] (193:1.45) -- ++(-1.5,0);
    \draw [dashed] (193:1.45) arc (80:50:1);
    
    \draw [->, >=stealth, scale=0.85] (A) ++ (0.15, 0.5) -- ++ (0,-0.75);
    \draw [->, >={stealth[bend]}, scale=0.85] (A)++ (0.15, 0.25) arc (180:255:0.25);
    \draw [->, >=stealth, scale=0.85] (A) ++ (-0.15, 0.5) -- ++ (0,-0.75);
    \draw [->, >={stealth[bend]}, scale=0.85] (A)++ (-0.15, 0.25) arc (0:-75:0.25);
    
    \draw [->, >=stealth, scale=0.85] (B) ++ (0.5, -0.15) -- ++ (-0.75,0);
    \draw [->, >={stealth[bend]}, scale=0.85] (B)++ (0.25, -0.15) arc (90:165:0.25);
    \draw [->, >=stealth, scale=0.85] (B) ++ (0.5, 0.15) -- ++ (-0.75,0);
    \draw [->, >={stealth[bend]}, scale=0.85] (B)++ (0.25, 0.15) arc (270:195:0.25);
    
    \draw [->, >=stealth, scale=0.85] (C) ++ (-0.15, -0.5) -- ++ (0,0.75);
    \draw [->, >={stealth[bend]}, scale=0.85] (C)++ (-0.15, -0.25) arc (0:75:0.25);
    \draw [->, >=stealth, scale=0.85] (C) ++ (0.15, -0.5) -- ++ (0,0.75);
    \draw [->, >={stealth[bend]}, scale=0.85] (C)++ (0.15, -0.25) arc (180:105:0.25);
    
    \draw [->, >=stealth, scale=0.85] (D) ++ (-0.5, 0.15) -- ++ (0.75,0);
    \draw [->, >={stealth[bend]}, scale=0.85] (D)++ (-0.25, 0.15) arc (270:345:0.25);
    \draw [->, >=stealth, scale=0.85] (D) ++ (-0.5, -0.15) -- ++ (0.75,0);
    \draw [->, >={stealth[bend]}, scale=0.85] (D)++ (-0.25, -0.15) arc (90:15:0.25);
    
    \draw [->, >=latex] (2.5,2) -- ++ (0,0.5) node[above, scale = 0.75] {N};
    \draw (2.3, 2.2) -- ++ (0.4,0);
 
\end{tikzpicture}
            \end{subfigure}\hfil
            \begin{subfigure}{0.5\textwidth}
            \centering
              \includegraphics[height=5cm, width=0.8\textwidth]{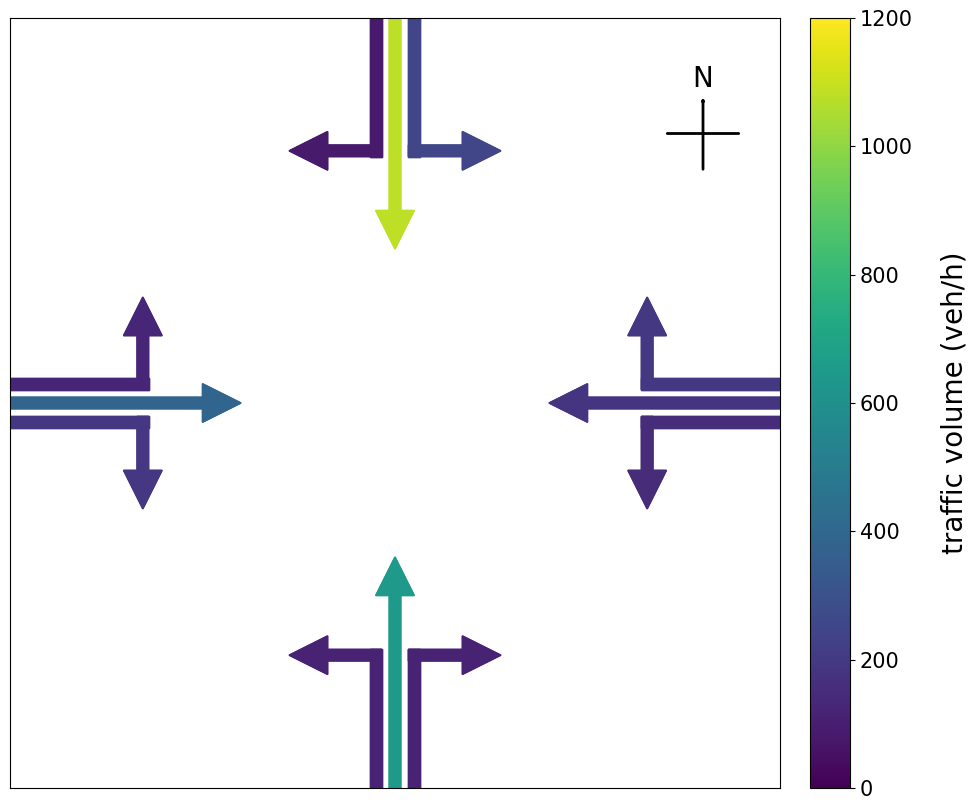}
            \end{subfigure}\hfil
          \caption{Illustration of roundabout capacity model. The required inputs are usually (left) roundabout geometry including permissible turns and (right) traffic volume by turn movement and vehicle class.}
          \label{fig:roundabout_modelling}
        \end{figure}
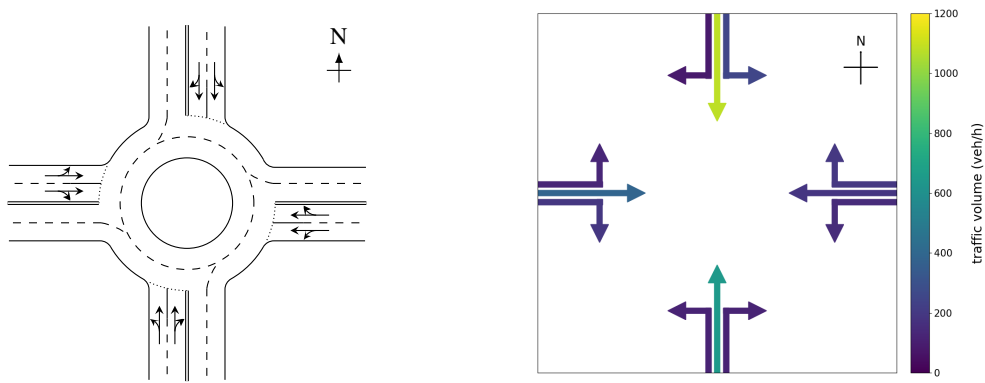

        A roundabout capacity model \citep{yap2013international} is an equations-based model that predict traffic performance measures at an intersection level. The capacity equations are either derived based on theory-driven results from queuing theory and gap acceptance process, data-driven empirical regression model, or a mixture of both. Given the intersection characteristics such as roundabout geometry and traffic volume as shown in Figure~\ref{fig:roundabout_modelling}, the usual task is to predict the aggregate traffic performance measures such as average capacity, delay and queue length. The parameters of interest are usually key gap acceptance parameters known to significantly affect the capacity, such as the critical gap $t_c$ defined as the minimum gap in the major stream traffic that a driver in the minor stream traffic is willing to accept \citep{troutbeck2014estimating}, and the followup headway $t_f$. Although such equations-based model is less versatile than simulation model described in the next example, it is often cheaper to evaluate and might be sufficient for the intended modelling task.

        \subsubsection*{Example 3: Network Simulation Model} \label{ex_3_network_simulation}
\begin{figure}[htbp!]
            \centering
            \includegraphics[height=8cm, width=\textwidth]{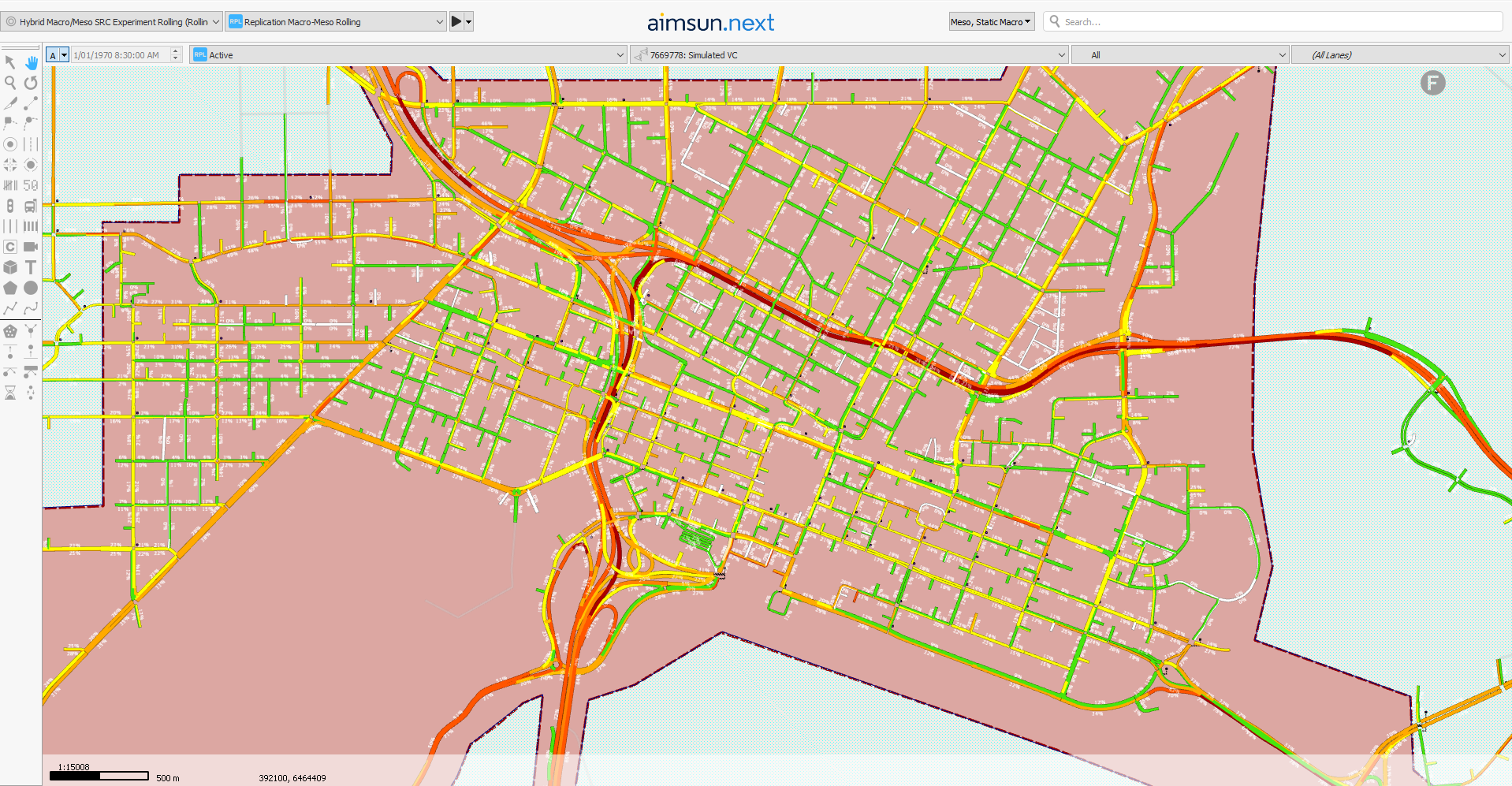}
            \caption{Illustration of network simulation model. This is a simulation for the CBD of Perth, Australia generated from Aimsun next \citep{AimsunManual}, a state-of-the-art traffic simulator.}
            \label{fig:network_simulation}
        \end{figure}

        Network simulation model as shown in Figure~\ref{fig:network_simulation}, often in the form of an agent-based model, is increasingly used to model large transport network with high fidelity. Since every agent is simulated, individual statistics can be recorded but usually only the summary statistics are of interest. Given the network characteristics such as network geometry, traffic control plan and traffic demand described by origin-destination (OD) matrices, the usual task is to predict the aggregate traffic performance measures such as average travel time, travel speed, delay etc. Although the parameters of interest are usually still driver behavioural parameters such as reaction time, each agent's reaction time is a realisation drawn from a probability distribution. Hence the parameters of interest are actually the parameters of the distribution such as the mean and variance. The network simulation model is stochastic up to random seed and its inference is fundamentally different from the previous two examples because the likelihood is not directly accessible (see section \ref{bayesian_computation}). 
        
    \subsection{Definitions and Notations}
    \sloppy
    This section formally defines all relevant terms for model calibration and validation. We first use the car following example to motivate the definitions and notations. In this case, the \emph{inputs} are the displacement, velocity and acceleration of both vehicles at time $t$, namely $[s_{n-1}(t), v_{n-1}(t), a_{n-1}(t), s_n(t), v_n(t), a_n(t)]^T$ and the \emph{outputs} are the displacement, velocity and acceleration of the follower at the next time step $t+\Delta t$, namely $[s_n(t + \Delta t), v_n(t + \Delta t), a_n(t + \Delta t)]^T$. Everything else that is required for the model to make a prediction, such as maximum desired acceleration, reaction time etc are the \emph{parameters}.

        \subsubsection{Inputs and Outputs} \label{input_and_output}
        Let the random vectors $X$ and $Y$ denote the \emph{inputs} and \emph{outputs} respectively. These random vectors have a joint distribution $(X, Y) \sim p(X, Y)$ that might be discrete, continuous or a mixture of both. In traffic modelling, $X$ is usually traffic demand and road geometry while $Y$ is usually traffic flow, capacity, density, speed, queue length, delay etc \citep{ni2004systematic}. Other names that one might encounter for $X$/$Y$ in the literature include: predictor/prediction, covariates/response, independent/dependent variables etc. The conceptual distinction between the two is that $X$ is specified by the modellers. Given a particular realisation $X=x$, the aim is to predict the likely realisations of $Y$. One such (point) prediction is its expected value $\mathbb{E}[Y|X=x]$.

        We call the joint distribution $p(X, Y)$ or\footnote{\label{note1}Which should be clear from the context.} the conditional distribution $P(Y|X)$ the \emph{reality} $R$. The usual assumption is that $X$ and $Y$ are \emph{correlated} so $p(Y|X) \neq p(Y)$ but it is not necessarily the case that $X$ \emph{causes} $Y$. Regardless of whether $R$ is inherently stochastic (aleatoric uncertainty) or the uncertainty is due to our lack of complete knowledge (epistemic uncertainty), we will inevitably model $R$ as a stochastic process\footnote{We will proceed in a Bayesian spirit of treating unknown quantities as random variables and describe them with the language of probability and statistics for the remainder of the paper. Even if a quantity is actually deterministic, it can be described by the Dirac measure so there is no loss in generality.}. The problem we face is that we never have complete knowledge of $R$ and any model (see section \ref{model}) that we construct is at best an approximation of it.

        Although $(X, Y)$ is often of main interest, we seldom observe their realisations $(x, y)$ directly but instead we observe their counterpart corrupted with measurement and estimation errors. Let $\tilde{X}$ and $\tilde{Y}$ denote the \emph{observed inputs} and \emph{observed outputs} respectively and viewed as random vectors. Then $(\tilde{X}, \tilde{Y}) \sim h(\cdot| X, Y, \sigma^2)$ where $h(\cdot)$ is the stochastic process of \emph{data collection} with \emph{noise parameters} $\sigma^2$. The difference between $(\tilde{X}, \tilde{Y})$ and $(X, Y)$ consists of \emph{measurement errors} due to limited instrumental precision as well as \emph{estimation errors}. The estimation errors arise because $(X, Y)$ might not be directly observable so we implicitly estimate them by constructing a statistic from other auxiliary variables that can be observed directly. In the network simulation example, $X$ might be the OD matrices between centroids which are not directly observed \citep{toledo2004statistical, toledo2004calibration} but instead estimated from detector counts across the network. Similarly in the car following example, $Y$ might be the follower's velocity time series which is not directly observed but instead calculated as the time derivative of the displacement time series (which is observed) with numerical differentiation and possibly smoothed with variants of Kalman filter \citep{punzo2005nonstationary}. These procedures introduce estimation errors when estimating the latent $(X, Y)$ and hence should be accounted for. We will refer to the joint distribution $p(\tilde{X}, \tilde{Y})$ or the conditional distribution $p(\tilde{Y}|\tilde{X})$ as the \emph{data generating process} (DGP). The DGP encapsulates the reality and the data collection process to result in the data that is ultimately observed.

        \subsubsection{Data} \label{data}
        We define a \emph{data point} $D_i = (\tilde{X}_i, \tilde{Y}_i)$ as a tuple of observed input and output indexed by $i$, and \emph{data} $\mathbf{D} = \{(\Tilde{X}_i, \tilde{Y}_i)\}_{i=1}^{N}$ as a set of $N$ such tuples where $(\tilde{X}_i, \tilde{Y}_i) \sim h(\cdot| X_i, Y_i, \sigma^2)$ and $(X_i, Y_i) {\sim} p(X_i, Y_i)$. Contrary to popular beliefs, independent and identically distributed (i.i.d.) data is not necessary for model calibration. The $D_i$'s need not be i.i.d. or even exchangeable (a weaker version of i.i.d., see section \ref{parameters}). In the car following example, suppose $X_i$ consists of the initial conditions, as well as the leader's displacement and velocity time series and  $Y_i$ is the follower's displacement and velocity time series. A car following model is typically calibrated using a longer time series $D_i = (X_i, Y_i)$ and validated on a different and shorter time series $D_j = (X_j, Y_j)$. As such, the random variables $D_i$ and $D_j$ are mapped to different measurable spaces with different dimensions and hence they cannot possibly be i.i.d.. Even if one defines a data point as the measurements at a specific time stamp so at least the codomain of each random variable is $\mathbb{R}$, autocorrelation still violates the independence assumption. However our intuition leads us to believe that learning suitable parameter values is still possible since the same vehicles are involved. We will elicit the actual assumptions in model calibration and validation in section~\ref{parameters} and the theorem that makes learning possible in principle.

        Since $\mathbf{D}$ is a collection of random vectors, it is a itself random vector. The literature often does not explicitly distinguish between the data as a random vector or a particular realisation because it is usually clear from the context. We will preserve this distinction as it is important for section~\ref{model_calibration} and call a particular realisation an \emph{observed data point} $d_i = (\tilde{x}_i, \tilde{y}_i)$, and \emph{observed data} $D = \{(\tilde{x}_i, \tilde{y}_i)\}_{i=1}^{N}$ respectively.
        
        \subsubsection{Model} \label{model}
        A model is often defined as an approximation to $R$ --- this is an oversimplification. We call $f(X, \mu, \xi)$ a \emph{simulator} which is a function of the inputs $X$, the \emph{calibration parameters} $\mu$ and possibly also $\xi$ denoting the stochastic element in the form of random seed. The simulator $f(\cdot)$ is a mathematical representation of $R$, often in the form of complex computer code\footnote{The word simulator is adopted from computer experiment literature \citep{bastos2009diagnostics}. A simulator is usually a deterministic function of all of its arguments, in the sense that evaluating it with the same arguments twice will return the same outputs. This however also includes stochastic models such as the network simulation example because the simulator is completely deterministic given the random seed that generates the sequence of pseudo random numbers.}. Its evaluation $f(x, \mu, \xi)$ at a given realisation of input $x$ and a given $\mu$ is an approximation of $p(Y|X=x)$ which is of main interest for modelling.
    
        Strictly speaking, the simulator $f(\cdot)$ alone is not the full model because it does not fully describe the DGP $p(\tilde{Y}|\tilde{X}=\tilde{x})$ and hence not directly related to the observed data. Indeed, \cite{ossen2008validity} showed that ignoring observation errors can yield considerable bias. Hence we define the \emph{model} $M(\cdot)$ as an approximation of the DGP as opposed to $R$. This implies $\tilde{Y} \mathrel{\dot\sim} M(\cdot|\tilde{X}=\tilde{x}, \theta)$ and the model $M(\cdot)$ consists of the following:
        \begin{align*}
            \hat{X} &\sim p(\hat{X}|\tilde{X} = \tilde{x}, \sigma^2_x)\\
            \hat{Y} &= f(\hat{X}, \mu, \xi)\\
            \Tilde{Y} &\sim p(\Tilde{Y}|\hat{Y}, \sigma^2_y)
        \end{align*}
        where $\hat{X}$ is an estimator of $X$ based on of $\tilde{X}$, $\hat{Y}$ is an estimator of $Y$ that depends on on $f(\cdot)$ and $\hat{X}$, $p(\cdot)$ is the observation model, $\sigma^2 = (\sigma_x^2, \sigma_y^2)$ are the observation noise parameters and $\theta = (\mu, \sigma^2)$ are the parameters.

\begin{figure}[htbp!]
            \centering
            \[
\tikz[
overlay]{
    \draw[dashed] (2.1, 1.2) rectangle (4.3,-1.2); 
}
\begin{tikzcd}
X \arrow{r}{h_x} \arrow[<->, swap]{d}{R} & \tilde{X} \arrow{r}{p_x} & \hat{X} \arrow{d}{f(\cdot, \mu)}\\
Y \arrow{r}{h_y} & \tilde{Y} & \hat{Y} \arrow[swap]{l}{p_y}
\end{tikzcd}
\]
            \caption{Notation relationships, $h(\cdot)$ and $p(\cdot)$ are data collection process and its approximation i.e. the observation model respectively. The model $M(\cdot)$ is the path in the dashed box.}
            \label{commutative_diagram}
        \end{figure}
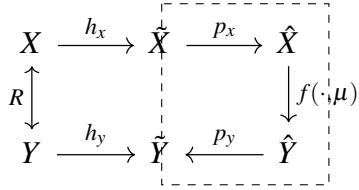

        As such, the simulator $f(\cdot)$ is usually only a part of the model and is equal to the model in the absence of the observation model. The relationships between the notations are summarised in Figure~\ref{commutative_diagram}. The statistical interpretation of model calibration is hence \emph{we seek a set of values for $\theta$ such that when given any realisation of observed inputs $\tilde{x}$, the conditional distribution $M(\cdot|\tilde{x}, \theta)$ can plausibly generate the corresponding $\tilde{y}$ as a realisation}.

        \subsubsection{Parameters} \label{parameters}
        Since the simulator takes both $\tilde{X}$ and $\mu$ as arguments, a key question that arises is indeed what distinguishes the (observed) inputs and the calibration parameters. An intuitive distinction based on observability, where $\mu$ is often thought of as non-observable quantities that need to be estimated, is inappropriate because $\tilde{X}$ is often also estimated in some sense, see section~\ref{data}. We claim that \emph{the modelling aim} is what decides whether a variable should be classified as an input or a parameter. In the roundabout capacity model example, the usual task treats traffic volume as $\tilde{X}$ and $t_c$ and $t_f$ as $\mu$. If the aim is to investigate the effect of population growth assuming the same driver characteristics, then such choices of $\tilde{X}$ and $\mu$ are perfectly sensible. However if the aim is to investigate the effect of a more aggressive driving behaviour given the same traffic volume, then the choices should be the other way. This illustrates that the inputs and parameters are not defined by the model architecture but rather by the modelling aim.

        Although model calibration usually focuses on estimating just $\mu$, we refer to $\theta$ as parameters for generality. The noise parameters $\sigma^2$ that form the remaining part of $\theta$ are nuisance parameters that also need to be accounted for. In classical statistics, the parameters are unknown constants which implies one cannot make probabilistic statements about them. In this paper, we will adopt a Bayesian paradigm (see section~\ref{bayesian_statistics_background} for more details) which allows one to view $\theta$ as a random vector with uncertainty due to a lack of information.
        
        We define \emph{parameters} as all the remaining arguments to $M(\cdot)$ aside from $\tilde{X}$. This is irrespective of whether their values are known precisely and hence there is no need for model calibration. In the roundabout capacity example, specifying the roundabout geometry requires information such as lane width, which can be considered as a parameter to be estimated. Despite the ability to measure lane width to high precision, this proposal is not as absurd as it sounds because of \emph{model imperfection error}. Since the simulator $f(X, \mu, \xi)$ is only an approximation of the reality $p(Y|X)$, the concept of lane width in reality as we understand does not necessarily translate to its counterpart in $f(\cdot)$ so the measured value is often not the best fitting value \citep{kennedy2001bayesian}. While some studies \citep{park2003microscopic, bayarri2004assessing, dowling2004guidelines, hollander2008principles} further distinguish between the parameters that one intends to estimate versus parameters that will be conditioned on specific values, we claim that the unifying perspective is \emph{model credibility}. If one believes $f(\cdot)$ is a faithful representation of $p(Y|X)$ then it would be sensible to treat observable variables such as lane width as known provided the measurement error is suitably low, otherwise it is more appropriate to treat it as an unknown quantity to be estimated.\\

        \paragraph{Exchangeability of Parameters \\}
        
        As alluded to in section~\ref{data}, the assumption in model calibration and validation is not necessarily that the $D_i$'s are i.i.d., but instead the parameters and/or the outputs are infinitely exchangeable given the inputs. A finite collection of random variables $Z_1, Z_2, ... , Z_n$ is said to be \emph{exchangeable} if the joint density is invariant under any permutation $(\pi_1, \pi_2, ... , \pi_n)$ of the indices $\{1, 2, ... , n\}$, that is 
        \[
        p_{Z_1, Z_2, ... , Z_n}(z_1, z_2, ... , z_n) = p_{Z_{\pi_1}, Z_{\pi_2}, ... , Z_{\pi_n}}(z_1, z_2, ... , z_n).
        \]
        A sequence of random variables $\{Z_i\}_{i=1}^{\infty}$ is \emph{infinitely exchangeable} if any finite collection $\{Z_i\}_{i=1}^{n}$ is exchangeable. Then by de Finnetti's representation theorem or its generalisations \citep{hewitt1955symmetric, diaconis1980finite}, for any $n$, there exists a parametric model $p(\cdot|\psi)$ indexed by some parameter $\psi$ such that 
        \[
        p(Z_1, Z_2, ... , Z_n) = \int \prod_{i=1}^{n} p(Z_i|\psi) p(\psi) d\psi.
        \]
        In other words, this means if we assume a sequence of random variables is infinitely exchangeable, then there must exist a common parameter $\psi$ and more importantly the random variables $Z_1, Z_2, ... , Z_n$ are \emph{conditionally i.i.d.} given $\psi$.
    
        If we assume that we can model each data point $D_i$ with its individual parameter $\theta_i$ and the sequence $\{\theta_i\}_{i=1}^{\infty}$ of parameters for all the data points that we could have observed is infinitely exchangeable. Then for the finite set of data that we did observe and their parameters $\{\theta_i\}_{i=1}^{N}$, the result above implies the existence of some latent hyperparameters $\psi$ (with assumptions on the functional form of priors) where
        \begin{align*}
            \tilde{Y}_i|\theta_i, \Tilde{X}_i &\mathrel{\dot\sim} M(\tilde{Y}_i|\tilde{X}_i, \theta_i)\\
            \theta_i|\psi &\sim p(\theta_i|\psi) \\
            \psi &\sim p(\psi)
        \end{align*}
        for $i= 1, 2, ... , N$ and the $\theta_i$'s are conditionally i.i.d. given $\psi$. Model calibration is concerned with learning parameters values, that is estimating the value of $\theta_i$ from $D_i$ and expecting this estimated value to be suitable for modelling $D_j$. Such learning is possible because the exchangeability assumption implies that the distributions of all $\theta_i$'s are parametrised by a common $\psi$. The plausible values of $\theta_i$ supported by observing $D_i$ constrains the plausible values of $\psi$, which then reduces our uncertainty about $\theta_j \sim p(\theta_j | \psi)$. Indeed, the assumption of i.i.d. $D_i$ common in machine learning is a stronger condition that implies the existence of a common $\theta$ so learning is simply estimating such $\theta$. In the more general yet common model calibration and validation settings, we often cannot assume $D_i$ to be i.i.d. or exchangeable (see section~\ref{data}). The common practice of obtaining a point estimate $\hat{\theta}$ from training data and subsequently checking if $\hat{\theta}$ is suitable for predicting validation data (see section~\ref{model_calibration} and \ref{model_validation}), implicitly further assumes that  $p(\cdot|\psi)$ can be well approximated by a Dirac delta function.
        
        Before ending this section, we emphasise the importance of \emph{parametrisation}. In the roundabout capacity example, suppose $X$ is traffic volume by turn movement\footnote{That is from which legs vehicles enter and exit. A turn movement is more formally defined as a tuple of approach leg and exit leg.}, and $\theta$ is the $t_c$ by turn type i.e. right turn, through, left turn etc. Now suppose we obtained data from two sites, one is a four-legged roundabout and the other a three-legged roundabout shaped like a T-junction. The sample spaces do not match but one might still reasonably expect the transferrability of the corresponding $t_c$ as per research carried out by \cite{gagnon2008calibration} and \cite{gallelli2017investigating}. Such intuition is in fact an assumption on the exchangeability of the $t_c$ for different sites. Even before data collection and model calibration, the parametrisation i.e. choices of calibration parameters should at least plausibly support the exchangeaibility assumption for one to expect model calibration to be useful and estimated parameters values to be transferrable.


\section{Model Calibration} \label{model_calibration}
We will now formalise the model calibration and validation problem in this and the following section. In particularly, we seek to answer the following three questions: 1) how does one obtain an estimate of $\theta$, 2) what does a good estimate mean and 3) how do we know if our estimate is good. This section addresses the first two questions which are closely related to model calibration. We first formulate the model calibration task using language from statistical inference and show where the commonly adopted optimisation based approach fits in as a specific case. We then provide a suitable definition of optimality of an estimate and illustrate how an optimisation based approach can be justified as a finite sample estimator. Finally, we summarise the current transport model calibration practices in the literature and point out the challenges.

    \subsection{Two Perspectives On Model Calibration} \label{alternatives}
    Our literature review discovered two major distinct approaches to model calibration. The first and the predominant approach is based on \emph{simulated minimum distance} \citep{grazzini2015estimation} which formulates model calibration as solving a certain optimisation problem  (see section~\ref{optimisation_based_calibration} for a detailed review) that involves comparing the model output and the observed output. We will synonymously refer to it as the optimisation based approach. The second approach is based on what we call \emph{physically meaningful calibration} where the parameters values are estimated directly in accordance to its assigned physical meaning. Direct estimation here means that the estimated parameters values is a function of only the observed data $D$ and do not require evaluations of the simulator $f(\cdot)$.
    
    In the roundabout capacity example, although the parameter $t_c$ is not directly observable, numerous statistical estimation methods based on the empirical distributions of observed accepted and rejected gaps were proposed to estimate it. For instance, \cite{troutbeck2014estimating} proposed the maximum likelihood estimator method and \cite{wu2012equilibrium} proposed probability equilibrium method. Other examples of such physically meaningful calibration approach can be found in \cite{treiber1999derivation, vasconcelos2014calibration, lu2016video} and \cite{durrani2019calibration}. 

    Despite their fundamentally different philosophy towards model calibration, the two approaches are not mutually exclusive. \cite{lu2016video} in fact combined both approaches in their model calibration task, where physically meaningful calibration is performed whenever the parameter is directly observable, and the optimisation based approach is only applied otherwise. 

    \cite{vasconcelos2014calibration} conducted a comparison study between the two approaches when calibrating Gipps' car following model. Their results suggest that the optimisation based approach offers good fit but poor transferrability while the physically meaningful calibration approach provides a satisfactory fit for a wide variety of traffic conditions. We will discuss the reasons and implications in section~\ref{optimality}. 

    Although the majority of the literature adopts the optimisation based approach as we will see in section~\ref{optimisation_based_calibration}, we emphasise that both approaches are sensible for model calibration with relative merits depending on model credibility as discussed in section~\ref{parameters}. In fact, it highlights a major difference between model calibration in engineering and model training in machine learning. The former is often explanatory with physically meaningful parameters while the latter focuses on flexibility in fitting arbitrary observed data at the expense of losing parameters interpretability.

    \subsection{Model Calibration As Estimators}
    The two seemingly disparate school of thoughts in section~\ref{alternatives} motivated a more general formulation of model calibration using ideas from statistical inference. Recall a \emph{statistic} $T(\mathbf{D})$ is a function of the data and hence is itself a random variable. An \emph{estimator} $\hat{\theta}$ is a statistic that is used to estimate the (population) parameters of interest namely $\theta$. An \emph{estimate} is the image of an estimator evaluated at a particular observed data $D$. In other words, an estimator is the procedure while an estimate is the actual values returned. Such distinction is important because we are concerned with variability which only makes sense when one speaks of an estimator. As such, \emph{model calibration} is more appropriately defined as \emph{the process of constructing an estimator $\hat{\theta}(\mathbf{D})$ and evaluating the corresponding estimate $\hat{\theta}(D)$}. We implicitly conditioned on specific model $M(\cdot)$ and typically we seek a \emph{point estimator}.

    The optimisation based approach can then be viewed as constructing an estimator of a specific form. Let $\theta$ denote a $p$-dimensional parameter vector and let $\Theta \subseteq \mathbb{R}^p$ denote the feasible parameters space. Let $C(\cdot, \cdot)$ denote a chosen \emph{cost function} which abstracts the notion of distance between the model output and the observed output. More formally, suppose dim($\tilde{Y}$) = $n_y$, then $C: \mathbb{R}^{n_y} \times \mathbb{R}^{n_y} \rightarrow \mathbb{R}^m$ where $m$ is often one for the common scalar cost function. A common name in transport model calibration literature for $Y$ and $C$ are \emph{measure of performance} and \emph{goodness-of-fit function} respectively but these are not as widely used in the statistics and machine learning community. Furthermore, let $\mathscr{A}_{\phi}(\zeta)$ denote an \emph{optimisation algorithm} belonging to some (possibly uncountable) family indexed by the hyperparameter $\phi$ and also taking another random vector $\zeta$ as argument. For instance, almost all optimisation algorithms contain some random elements and is iterative in nature, even deterministic direct search will often have random initialisation. In practice, both $\xi$ and $\zeta$  will be the random seed of the underlying pseudo-random number generator. The optimisation based approach adopted in most model calibration studies essentially constructs an estimator that takes the form of an \emph{extremum estimator}\footnote{Strictly speaking, an extremum extimator is an estimator that is a solution to an optimisation problem without consideration of the Monte Carlo error due to imperfect optimisation algorithm. We will avoid introducing redundant and pedantic notations. In this paper, the optimisation algorithm is always considered for practical application.} as follows:
    \begin{align*}
    \hat{\theta}(D)|M &= T(D) | M, C, \mathscr{A} \\
                    &= \argmin_{\theta \in \Theta} \: C(M(\Tilde{X}, \theta), \Tilde{Y}) \Big|_{\mathscr{A}}
    \end{align*}
    where we have suppressed the appropriate expectations for now for clarity, but are technically needed to ensure the problem is well-posed. The conditioning bar denotes that $\mathscr{A}$ is the solver for the optimisation problem. One interpretation is that $\mathscr{A}$ induces a stochastic process on the state space $\Theta$. We make the dependence of the estimator on $C$ and $\mathscr{A}$ explicit because despite them being artefact of calibration choices, they will influence how good an estimate is. 
    
    The most popular point estimator from statistical inference, namely the maximum likelihood estimator which is adopted in \cite{hoogendoorn2010generic} and \cite{xu2020statistical}, is clearly a specific case of the extremum estimator where $C(\cdot, \cdot)$ is the negative log-likelihood. Although extremum estimator comprises a broad class of estimators and indeed a lot of popular estimators can be phrased in terms of an optimisation problem\footnote{\label{note2} For instance, the posterior mean $\bar{\theta} = \int \theta p(\theta|D) d\theta$ is the estimator that minimises the expected mean squared error, i.e. $\bar{\theta} = \argmin_{\theta'} \mathbb{E}_{\theta|D}[(\theta' - \theta)^2]$ where the expectation is taken over the posterior distribution of $\theta$.}, it is certainly not the only approach to constructing an estimator. The formulation of model calibration in terms of estimators is sufficiently general and unifies the two perspectives in section~\ref{alternatives}.
    
    \subsection{Definition of Optimality} \label{optimality}
    The role of model calibration is hence effectively a choice of an appropriate estimator. In order to assess how good an estimator is, we first need to define what optimality of an estimator means. Since $M(\cdot)$ is only an approximation of the DGP, there is no \emph{true parameter value} even if $\theta$ can be sensibly assigned a physical meaning. One can however sensibly define an optimal parameter value as \emph{what is most useful for predictions}. As such, we define the \emph{optimal parameter value} as one that minimises the \emph{generalised error}:
    \begin{align}
    \theta^*|M, C := \argmin_{\theta} \:\: \mathbb{E}_{\mathbf{D}} [ C(M(\tilde{X}, \theta), \tilde{Y})]. \label{optimal_param_def}
    \end{align}
    We make the dependence on $M$ and $C$ explicit in \eqref{optimal_param_def} to emphasise that optimality is only defined w.r.t. to choices of them. In other words, $\theta^*$ minimises the expected difference between simulated output and observed output where the expectation is taken over the data distribution. If $M(\cdot)$ involves a stochastic simulator $f(\cdot)$ with randomness due to $\xi$, then an inner expectation of $\xi$ is also required. The expectation in (\ref{optimal_param_def}) cannot be computed because the DGP which describes the data distribution is unknown. However this definition is fundamental to imbue meaning to model calibration. In the remainder of this paper, we will assume such $\theta^*$ exists.\footnote{Some authors distinguish between parameters estimation and model calibration, namely the assumption that a true value exists in estimation while calibration merely finds a best fitting value knowing the model is imperfect. However the distinction is not well-defined \citep{grazzini2015estimation} and we see model calibration as conditional parameters estimation where the role of the true value is replaced by the best fitting value.}

    The optimality of a model calibration methodology, or equivalently the optimality of the corresponding estimator is commonly measured by its mean squared error (MSE) to $\theta^*$ in statistical inference, namely:
    \begin{align*}
    \text{MSE}(\hat{\theta}, \theta^*) &= \mathbb{E}[(\hat{\theta} - \theta^*)^2] \\
                             &= \mathbb{E}[(\hat{\theta} - \mathbb{E}[\hat{\theta}] + \mathbb{E}[\hat{\theta}] - \theta^*)^2] \\
                             &= \mathbb{E}[(\hat{\theta} - \mathbb{E}[\hat{\theta}])^2] + 2\mathbb{E}[(\hat{\theta} - \mathbb{E}[\hat{\theta}])(\mathbb{E}[\hat{\theta}] - \theta^*)] + (\mathbb{E}[\hat{\theta}] - \theta^*)^2 \\
                             &= \mathbb{E}[(\hat{\theta} - \mathbb{E} [\hat{\theta}])^2] + (\mathbb{E}[\hat{\theta}] - \theta^*)^2
    \end{align*}
    where the expectation is taken over the distribution of $\hat{\theta}$. The MSE allows simple derivation of arguably the most important concept in statistical learning which is foundational to model calibration and validation, namely the \emph{bias-variance tradeoff}. The first term in the last line $\mathbb{E}[(\hat{\theta} - \mathbb{E} [\hat{\theta}])^2]$ is the variance of $\hat{\theta}$ and the second term $(\mathbb{E}[\hat{\theta}] - \theta^*)^2$ is the (square of) bias. In section~\ref{bias_variance_tradeoff}, we will discuss the profound implications of bias-variance tradeoff such as over(under)-fitting which necessitates model validation.
    
    The typical questions from estimation theory are concerned with properties of estimators. For instance: is $\hat{\theta}$ unbiased, i.e. is $\mathbb{E}[\hat{\theta}] = \theta^* \: \forall \theta^* \in \Theta$? Is $\hat{\theta}$ efficient\footnote{Efficiency clearly depends on the chosen loss function to which MSE is only one such choice. One might ask if there exists the most efficient estimator which could be a promising candidate in model calibration. A relevant direction concerns what is called the \emph{Cramer-Rao Lower Bound} (CRLB) which is a lower bound on the variance of an unbiased estimator and the uniformly minimum variance unbiased estimators (UMVUE) which are unbiased estimators whose variance is less than or equal to that of all unbiased estimators, irrespective of whether CRLB is achievable. We refer readers to more specialised text such as \cite{abramovich2022statistical} for details.}, i.e. is $\text{MSE}(\hat{\theta}, \theta^*) < \text{MSE}(\hat{\theta}', \theta^*)$ for any other estimator $\hat{\theta}'$ ? Is $\hat{\theta}$ asymptotically consistent\footnote{An extremum estimator is consistent for a global optimiser under certain conditions, see \cite{newey1994large}} w.r.t sample size so $\hat{\theta}$ converges to $\theta^*$ in probability as more data is gathered? More formally, let $\hat{\theta}_n$ denote an estimator with data of size $n$, then does $\lim_{n \rightarrow \infty} P(||\hat{\theta}_n - \theta^*|| < \epsilon) = 1 \: \forall \epsilon > 0$?

    If the quality of an estimator is assessed in terms of MSE, then unbiasedness is merely a desirable property but not necessarily optimal as introducing some bias can cause significantly more variance reduction. The asymptotic consistency property is indeed crucial for questions such as ``does more data help?'' which is often taken for granted. Unfortunately, these important theoretical considerations are often left inconclusive due to the complexity and black box nature of $f(\cdot)$ where even crucial regularity condition such as continuity cannot be guaranteed.
    
    We end this section with a reminder of the \emph{standard error} which is the standard deviation of $\hat{\theta}$ and describes the sampling variability. Even if one can marginalise out the stochastic elements of $M(\cdot)$, there is inherent variability in $\hat{\theta}$ based on different realisations of data that we could have observed. Based on our review of traffic model calibration studies, only \cite{hoogendoorn2010generic} investigated this sources of variability. Most calibration studies are instead exclusively concerned with variability due to a stochastic $f(\cdot)$ or $\mathscr{A}$ which are in fact sources of \emph{Monte Carlo error}. The likely reason is the prohibitively expensive collection of large amount of data required. Model calibration in fact needs to address both sources of variability, a point we will return to in section~\ref{numerical_optimisation}.
    
    \subsection{A Review of Optimisation-Based Framework} \label{optimisation_based_calibration}
    Due to their predominance, we will provide an in-depth review of the optimisation based model calibration frameworks in the literature. We will also point out some subtleties and potential confusion at each step.
    
        \subsubsection{Parameters Selection}
        Assuming $M(\cdot)$ is already decided, the first step in all model calibration and validation problems, even if only implicitly alluded to, is choosing problem specific parametrisation. For instance, what is $D$ and how should it be collected with an acceptable observation error? What are $\tilde{X}, \tilde{Y}$  and $\theta$? In particular, modern simulator allows modelling of of traffic problems with large number of variables and increasing complexity. The choices of inputs, outputs and parameters are not merely conventional but instead depends on the modelling aim, see section~\ref{parameters}.

        While $M(\cdot)$ often takes many arguments and all of them except those designated as $\tilde{X}$ are technically parameters, model calibration is often only concerned with estimating a small subset of them both for identifiability (see section~\ref{practical_considerations}) and computational feasibility. This section focuses on the task of \emph{parameters selection}. Let $\tilde{\theta} \in \Tilde{\Theta} \subseteq \mathbb{R}^P$ denote the full $P-$dimensional parameter vector. WLOG, the parameters selection task is equivalent to choosing a parametrisation such that $\tilde{\theta} = [\theta, \theta^{c}]^T \in \Theta \times \{0\}^{P-p}$ where $\theta \in \Theta \subseteq \mathbb{R}^p$ and $p << P$. In other words, only a much lower dimensional parameter $\theta$ will be estimated and the remaining parameters $\theta^c$ will be conditioned on some pre-defined values.

        For a black box $f(\cdot)$, modellers often lack prior knowledge on parameters interaction and correlation which are important \citep{kim2011correlated}. The common practices in the literature are\footnote{We will avoid generic advice such as 'incorporate domain expertise' which is universally applicable but provides little guidance.}:
        \begin{itemize}
            \item \emph{Sensitivity Analysis}\\
            Sensitivity analysis such as the simplest one-at-at-time \citep{lownes2006sensitivity}, variance-based decomposition\footnote{Aside from computational complexity, one-at-a-time and variance-based decomposition also differ in terms of local versus global sensitivities.} \citep{ciuffo2013gaussian} or analysis of variance \citep{park2005development} estimates how sensitive the model output is to variation in different parameters. The parameter to which the model output is most sensitive to is supposedly important and hence chosen.
            \item \emph{Dimension Reduction Techniques}\\
            Dimension reduction techniques project to lower dimension while trying to retain as much information as possible. The most common approach is principle component analysis which finds the linear combination of variables that preserves the most variance. For parameters that have a hierarchical structure, a clustering approach can be adopted where lower level parameters are assigned a common value. In the network simulation example, the road network often spans large geographical region and contains many different road types such as freeway, arterial roads, local roads etc. Suppose free flow speed for each road segment is a parameter, then a clustering approach might sensibly assign the same free flow speed to all road segments that are local roads.
            \item \emph{Default Values}\\
            A theory-driven $f(\cdot)$ might supply \emph{default parameter value} which is in fact an estimator constructed based on prior knowledge. When $\theta^c$ is excluded from model calibration, it is effectively assigned some constant value that is often the default value. Arguably, a modeller should start with prior predictive check using the default values to first justify the necessity of model calibration before proceeding.
        \end{itemize}

        We end our discussion on parameters selection by addressing \emph{conditional parameters} which are parameters whose existence depends on other settings. In the roundabout capacity example, suppose we na\"ively chose $\theta$ as $t_c$ by turn movement. The dimension of $\theta$ changes for different roundabouts with different turn movement configurations and there is no clear correspondence between say the North-West movement of two sites. The issue is not just shape incompatibility during computation but rather the more subtle conceptual flaw that the exchangeability assumption in section~\ref{parameters} does not hold. Hence one cannot expect the estimates to be transferable and alternative parametrisation might be necessary.
        
        \subsubsection{Numerical Optimisation} \label{numerical_optimisation}
        A combination of $\mathbf{D}, \theta, C, \mathscr{A}$ results in a specific estimator. The objective function and hence the corresponding optimisation problem commonly stated in the methodology section of most literature is in fact such estimator evaluated at the observed $D$. The optimisation problem is expressed below in a more conventional and explicit form:
        \begin{align}
        \argmin_{\theta} \quad & \biggl[ \frac{1}{|D|}\sum_{(\tilde{x}_i, \tilde{y}_i) \in D} \frac{1}{|\Xi_i|}\sum_{\xi_{ij} \in \Xi_i} C_i( M_{\xi_{ij}}(\tilde{x}_{i}, \theta), \: \tilde{y}_{i}) \biggr] \quad + \quad \lambda R(\theta) \label{loss_function}\\ 
        \textrm{s.t.} \quad & \theta_{(j)} \in [L_j, \: U_j], \quad 1 \leq j \leq n_c \label{constraint cv}\\
                        & \theta_{(j)} \in \{c_{j_{1}}, \: c_{j_{2}}, \: ... \:,  c_{j_{N_j}}\}, \quad n_c+1 \leq j \leq n_c + n_d \label{constraint dv}\\
                        & G_k(\theta) = 0, \quad 1 \leq k \leq K_{1} \label{equality}\\ 
                        & H_k(\theta) \leq 0, \quad 1 \leq k \leq K_{2} \label{inequality}
        \end{align}
        WLOG the $n_c$ continuous variables are ordered before the $n_d$ discrete variables so $\theta \in \mathbb{R}^{n_c} \times \mathbb{Z}^{n_d}$ where $n_c + n_d = p$. Equations~\eqref{constraint cv}-\eqref{constraint dv} describe the typically bounded parameter space while \eqref{equality}-\eqref{inequality} are the possible additional $K_1$ equality and $K_2$ inequality constraints that a feasible solution also needs to satisfy respectively\footnote{This is somewhat redundant and included for completeness because any equality constraints $G_i(\theta) = 0$ is equivalent to the two inequality constraints $G_i(\theta) \leq 0$ and $G_i(\theta) \geq 0$.}. Clearly, \eqref{constraint cv}-\eqref{inequality} collectively define $\Theta$.
        
        We will refer to the objective function in \eqref{loss_function} as the \emph{loss function}. The first term in the loss function is the most common form of $C$ and is usually referred to as the residual. The $C_i(\cdot,\cdot)$'s are the point-wise cost function that forms part of $C$ but $C$ does not necessarily decompose into such additive form especially if $D_i$'s are not i.i.d. (see section~\ref{data}). The outer sum averages over the observed data points which only makes sense for i.i.d. data and the inner sum averages over multiple runs of the stochastic model indexed by $\xi_{i,j}$ which concern variability due to standard error and Monte Carlo error respectively. If optimality is defined according to (\ref{optimal_param_def}), then an optimisation based approach is essentially constructing a Monte Carlo approximation of the generalised error $\mathbb{E}_{\mathbf{D}} [ C(M(\tilde{X}, \theta), \tilde{Y})]$ with finite sample and finite number of model runs. The second term in the loss function $\lambda R(\theta)$ is referred to as the \emph{regulariser}. Its role is to increase the well-posedness of loss function because simply minimising the residual can easily lead to over-fitting (see section~\ref{bias_variance_tradeoff}). For instance, \cite{toledo2004calibration} and \cite{ossen2005car} penalised deviation from apriori assumed optimal values.

        The challenges of numerically solving \eqref{loss_function} is usually discussed in the context of \emph{global optimisation} and/or \emph{black-box optimisation}. The dynamic model $f(\cdot)$ and hence the loss function is often non-linear and non-convex\footnote{Perhaps even non-differentiable or non-continuous at some points such as with piece-wise definition. Although one might question the physical reasonableness of such discontinuity, it might still be a reasonable approximation to model say phase transitions.} so most calibration studies utilise some optimisation algorithm $\mathscr{A}$, often stochastic global optimiser. We reiterate the somewhat obvious facts that can still cause confusion in existing literature if not phrased carefully: 1) the global minimum of \eqref{loss_function} might not exist if the loss function is not continuous w.r.t. $\theta$ or $\Theta$ is not compact, 2) even if it exists, the global minimum might not be unique if the loss function is non-convex w.r.t. $\theta$ and 3) even if the global minimum exists and is unique, there is often no guarantee that $\mathscr{A}$ converges to it within sufficient tolerance in finite time. Without proof, it is incorrect to claim that one has found the global optimum just because the optimisation algorithm converges to something. At best one can only hope for convergence to a sufficiently good local optimum as the empirical solution of \eqref{loss_function}.
        
       The literature on traffic model calibration methodology usually focuses on the choice of $C$ and $\mathscr{A}$. Table~\ref{tab:cost_function_and_optimiser} is a summary of the common choices in existing studies which is by no means exhaustive. Due to space constraints, readers are also referred to other reviews such as \cite{hollander2008principles, yu2017calibration, sha2020applying} and \cite{punzo2021calibration}.\\

\begin{sidewaystable}[htbp!]
    \scriptsize
    \setlength{\tabcolsep}{2pt}
    \renewcommand\cellgape{\Gape[2pt]}
    \centering

        \begin{threeparttable}
        \captionsetup{font=scriptsize}
        \caption{Combinations of $C$ and $\mathscr{A}$ used in the literature. Studies without explicit mention of the optimisation algorithm used were excluded. Other notable cost functions include: (the p-value of) Kolmogorov-Smirnov Test Statistic $\max |F_x - F_y|$ \citep{kim2005calibration}. Other notable optimisation algorithms include: simulated annealing \citep{ciuffo2013no}, tabu search \cite{yu2017calibration}, particle swarm \cite{karimi2019two} and Bayesian optimisation \cite{sha2020applying}.}
        \label{tab:cost_function_and_optimiser}
        
        \begin{tabular}{|l||*{6}{p{3.8cm}|}}\hline
        \diagbox[width=15em]{$C$}{$\mathscr{A}$}
        &\makecell[c]{Simplex}
        &\makecell[c]{GA}
        &\makecell[c]{SPSA}
        &\makecell[c]{NLP\tnote{1}}
        &\makecell[c]{Manual\tnote{2}}
        \\
        \hline
        \hline
            \makecell{Root Mean Square Error \\ (RMSE) \\ $\sqrt{\frac{1}{N} \sum_{i=1}^N (x^{\text{sim}}_i - x^{\text{obs}}_i)^2}$} 
            &\makecell{\cite{toledo2004calibration}\tnote{a, *, \textdagger, \textdaggerdbl} \\ \cite{punzo2012can}\tnote{d} }
            &\makecell{\cite{kesting2008calibrating}\tnote{d, \textdagger} \\ \cite{punzo2012can}\tnote{d} \\ \cite{ciuffo2013no}\tnote{b, \textdagger} \\ \cite{vasconcelos2014calibration}\tnote{d} \\ \cite{giuffre2018calibrating}\tnote{c, \textdagger} \\ \cite{chiappone2016traffic}\tnote{b, *, \textdagger} \\ \cite{punzo2021calibration}\tnote{d, *, \textdagger} }
            &\cite{ciuffo2013no}\tnote{b, \textdagger, $\S$}
            &\makecell{\cite{hourdakis2003practical}\tnote{b, \textdagger} \\ \cite{ciuffo2008comparison}\tnote{b} \\ \cite{punzo2012can}\tnote{d} \\\cite{ciuffo2013no}\tnote{b, \textdagger} }
            &
            \\
        \hline
            \makecell{Root Mean Square Percentage Error \\ (RMSPE) \\ $\sqrt{\frac{1}{N} \sum_{i=1}^N (\frac{x^{\text{sim}}_i - x^{\text{obs}}_i}{x^{\text{obs}}_i })^2}$}  
            &
            &\makecell{\cite{kesting2008calibrating}\tnote{d, \textdagger} \\ \cite{ciuffo2013no}\tnote{b} \\ \cite{gallelli2017investigating}\tnote{c} \\ \cite{punzo2021calibration}\tnote{d, *, \textdagger} }
            &\makecell{\cite{ciuffo2013no}\tnote{b, $\S$} \\ \cite{sha2020applying}\tnote{a, *} }
            &\makecell{\cite{ciuffo2008comparison}\tnote{b, *} \\ \cite{ciuffo2013no}\tnote{b} }
            &\makecell{\cite{hourdakis2003practical}\tnote{b}}
            \\
        \hline

            \makecell{Mean Absolute Error \\ (MAE) \\ $\frac{1}{N} \sum_{i=1}^N |x^{\text{sim}}_i - x^{\text{obs}}_i|$}  
            &\makecell{\cite{kim2003simplex}\tnote{b} \\ \cite{brockfeld2004calibration}\tnote{d, \textdagger} \\ \cite{punzo2012can}\tnote{d} }
            & \makecell{\cite{ma2002genetic}\tnote{a, \textdagger} \\ \cite{park2005development}\tnote{c, \textdagger} \\ \cite{park2006application}\tnote{a, \textdagger} \\ \cite{punzo2012can}\tnote{d} \\ \cite{ciuffo2013no}\tnote{b} \\ \cite{punzo2021calibration}\tnote{d} }
            &\cite{ciuffo2013no}\tnote{b, $\S$}
            &\makecell{\cite{punzo2012can}\tnote{d} \\ \cite{ciuffo2013no}\tnote{b}}
            &\\
        \hline
            \makecell{Mean Absolute Percentage Error \\ (MAPE) \\ $\frac{1}{N} \sum_{i=1}^N |\frac{x^{\text{sim}}_i - x^{\text{obs}}_i}{x^{\text{obs}}_i }|$}
            &
            &\makecell{\cite{ciuffo2013no}\tnote{b} \\ \cite{yu2017calibration}\tnote{b, *} \\ \cite{punzo2021calibration}\tnote{d, *} }
            &\makecell{\cite{ciuffo2013no}\tnote{b, $\S$} \\ \cite{hale2015optimization}\tnote{b, \textdagger} }
            &\cite{ciuffo2013no}\tnote{b}
            &\cite{hale2015optimization}\tnote{b, \textdagger}
            \\
        \hline
            GEH 
            &\makecell{\cite{punzo2012can}\tnote{d} }
            &\makecell{\cite{ma2007calibration}\tnote{b, *} \\ \cite{punzo2012can}\tnote{d} \\ \cite{ciuffo2013no}\tnote{b} }
            &\makecell{\cite{ma2007calibration}\tnote{b, *} \\ \cite{ciuffo2013no}\tnote{b, $\S$} }
            &\makecell{\cite{punzo2012can}\tnote{d} \\ \cite{ciuffo2013no}\tnote{b} }
            &
            \\
        \hline
            \makecell{Theil's U \\ $\frac{\sqrt{\frac{1}{n}\sum_{i=1}^{n}(x_i-y_i)^2}}{\sqrt{\frac{1}{n}\sum_{i=1}^{n}y_i^2} + \sqrt{\frac{1}{n}\sum_{i=1}^{n}x_i^2}}$}
            &\makecell{\cite{brockfeld2005calibration}\tnote{d} \\ \cite{ossen2008validity}\tnote{d} \\ \cite{ossen2009reliability}\tnote{d, *} \\ \cite{punzo2012can}\tnote{d} }
            &\makecell{\cite{punzo2012can}\tnote{d} \\ \cite{ciuffo2013no}\tnote{b}\\ \cite{punzo2021calibration}\tnote{d, *} }
            &\cite{ciuffo2013no}\tnote{b, $\S$}
            &\makecell{\cite{punzo2012can}\tnote{d} \\ \cite{ciuffo2013no}\tnote{b} }
            &\cite{hourdakis2003practical}\tnote{b}
            \\
        \hline

        \end{tabular}
        \begin{tablenotes}
            \item[*] Linear combination with different outputs
            \item[\textdagger] With alternative normalisation or monotonic transformation, which is equivalent to $C$ from an optimisation point of view for fixed data i.e. the solution is the same but it can affect the efficiency of the optimiser.
            \item[\textdaggerdbl] They used an extension of simplex called Box's Complex algorithm which is essentially a constrained simplex method.
            \item[\S]  They also used a second order SPSA.
            \item[1] These studies often utilise optimisation packages which uses reduced gradient methods for constrained nonlinear optimisation.
            \item[2] We consider this to include exchaustive grid search.
            \\
        \end{tablenotes}
        Legend for traffic modelling context
        \begin{tablenotes}
            \item[a] Network simulation
            \item[b] Freeway sections simulations
            \item[c] Intersection modelling
            \item[d] Car Following
        \end{tablenotes}
        \end{threeparttable}
\end{sidewaystable}

        \paragraph{Cost Function\\}

        The main conceptual difference between different $C$ is 1) whether larger errors are penalised much more and 2) whether different variables are normalised. Cost functions such as $RMSE$ penalises larger errors disproportionately more than smaller errors. Some authors advocate such property because a large error is more plausibly a clear sign of model misfit while a series of small errors might be more appropriately attributed to observation noise. The rationale of normalising in cost functions such as $RMSPE$ is to conduct some form of scaling so only relative errors are considered and different output components become comparable. 

        \paragraph{Optimisation Algorithm\\}

        If $p$ is low dimensional, a grid search over $\Theta$ might be feasible which allows one to visualise the optimisation landscape and hence a good approximation to the solution of \eqref{loss_function}. This is often not the case so one resorts to some optimisation algorithm $\mathscr{A}$ that we broadly classified into four categories as follows:
        \begin{itemize}
            \item Non-automated search\\
            Non-automated search methods are often referred to as manual adjustment or ad hoc adjustment. Although traditionally perceived as slow, subjective and time-consuming \citep{hourdakis2003practical} which motivated automated optimisation algorithms, it can achieve a close match by trained experts. Such subjective approach might also be advantageous when viewed as a method that can incorporate prior belief of the modellers and quickly eliminate values that are not physically plausible. 
            \item Direct search\\
            The direct search methods are derivative-free and only uses evaluations of the loss function. They progressively keep the best solution found so far and occasionally deviate as attempts to escape local optima. The advantage of direct search methods is their applicability to black box models and handling of discrete parameters but they are generally much slower to converge. Convergence here refers to termination of algorithm because the best solution found stops improving. As shown in Table~\ref{tab:cost_function_and_optimiser}, genetic algorithm (GA) is by far the most common.
            \item Gradient-based methods\\
            Gradient-based methods use the gradient of \eqref{loss_function} w.r.t. $\theta$ or higher order derivatives such as the Hessian to guide the search. They are generally much faster to converge but the gradient might not be readily available. Although numerical derivative using finite difference is one workaround, it is often prohibitively expensive because each iteration now requires $O(p)$ evaluations of $f(\cdot)$ and can be numerically unstable. The most common gradient-based method in traffic model calibration appears to be the simultaneous perturbation stochastic approximation (SPSA) which estimates the gradient using only two model evaluations by simultaneously perturbing every component of $\theta$ but such estimate can be noisy.
            \item Surrogate model approaches\\
            A surrogate model $\hat{f}(\cdot)$ is an approximation of the original simulator $f(\cdot)$ and is particularly useful when $f(\cdot)$ is expensive to evaluate. The idea is to first use a limited number of runs with $f(\cdot)$ to fit $\hat{f}(\cdot)$ so as to mimic the response of $f(\cdot)$. Once a sufficient fit is obtained, $\hat{f}(\cdot)$ is then used to guide the search by proposing promising candidate solution $\theta'$ to be assessed. The assessment evaluation $f(\theta')$ might be used to subsequently improve the surrogate fit. The most commonly used surrogate model is Gaussian process (GP) regression \citep{ciuffo2013no, ciuffo2013gaussian, sha2020applying} which in the optimisation context is also referred to as Bayesian Optimisation (BO). Aside from much cheaper function evaluation, the GP surrogate often allows closed form results to be derived and hence permits efficient use of higher order search methods such as L-BFGS \citep{liu1989limited}. It is worth mentioning that the surrogate model does not necessarily need to be a flexible and practically non-parametric model such as GP or neural network \citep{otkovic2013calibration}. Other attempts in the literature include \cite{park2003microscopic} who fitted a linear regression or more generally a lower fidelity approximation of $f(\cdot)$, see \cite{punzo2007steady, rakha2009procedure} and \cite{patwary2021metamodel}
        \end{itemize}

        For completeness, we also summarise below some heuristics and strategies that might be useful in solving \eqref{loss_function}. \cite{toledo2003calibration} and \cite{toledo2004calibration} used an iterative approach that switches between optimising for different subsets of parameters which in their study consist of estimating OD matrices and estimating driver behavioural parameters. \cite{hourdakis2003practical} and \cite{dowling2004guidelines} adopted a hierarchical approach where global parameters are calibrated before calibrating local parameters. Similarly, \cite{hoogendoorn2010generic} conducted joint estimation of the $\theta_i$'s and the parameters that describe their probability distribution. This is in fact a hierarchical modelling approach which is more naturally handled by Bayesian inference, see section~\ref{flexibility_and_statistical_coherence}. Lastly, \cite{yu2017calibration} proposed a warm start approach where the solution found from one optimisation algorithm is used as the starting point for another optimisation algorithm.

        A suitable choice of $C$ and $\mathscr{A}$ is undoubtedly context-specific and requires extensive comparison studies which is still an area that needs further research. We will refer readers to literature dedicated to such context specific comparisons such as \cite{kesting2008calibrating, punzo2012can, ciuffo2013no} and \cite{punzo2021calibration}. Due to space constraints, we also refer readers to a more comprehensive review on cost functions, see \cite{cha2007comprehensive} and on optimisation algorithms, see \cite{boussaid2013survey} and \cite{giagkiozis2015overview}. 
      
        \subsubsection{Practical Considerations} \label{practical_considerations}
        We will end our review on optimisation based model calibration by summarising the challenges during optimisation which are seldom discussed in the literature and the strategies adopted. An optimisation based approach to model calibration, similar to training in machine learning (ML), is primarily driven by monitoring the training error $\epsilon_{tr}$ which is the first term in \eqref{loss_function} with $D=D_{tr}$, the training data set. We will illustrate below why such error-centric inference procedure, which is tremendously successful in ML, might not fully applicable to model calibration.

        \paragraph{Achievable Error\\}
        A target error or even just an acceptable error threshold $\epsilon_{th}$ should technically be established prior to running the optimisation algorithm to be able to conclude sufficient model fit. However unlike ML models, traffic models often contained theoretically derived components which make them much more rigid. Even with a large number of parameters, they are rarely guaranteed to be \emph{universal approximators} so a desirable low threshold might simply not be achievable even with a perfect optimisation algorithm that can find the global optimum of \eqref{loss_function}. The strategy attempted in the literature to estimate such error threshold is to conduct some form of \emph{ensemble comparison} which compares the minimum error achieved when fitting the model to different data set and/or fitting several conceptually similar model (with ideally similar complexity) to the same data set \citep{brockfeld2003toward, brockfeld2004calibration, brockfeld2005calibration}.

        Even if $\epsilon_{th}$ is achievable, we ultimately face the practical issue of computational infeasibility. The usual black box treatment of $f(\cdot)$ often prohibits efficient optimiser such as backpropagation in training neural networks. As a result, Monte Carlo error due to variability in locating the good local optima across different optimisation runs in finite time can be significant. The most common approach to investigate the efficiency of an optimiser is to conduct \emph{benchmarking on synthetic data} \citep{ossen2008validity, ossen2009reliability, punzo2012can, ciuffo2013no}, where the model is used to generate the data and one seeks to recover the parameters values used. Such benchmarking can be used to compare different family of optimisation algorithms \citep{ma2007calibration, sha2020applying} or to search within one family a promising value of the hyperparameter $\phi$. However, synthetic data ultimately assumes a perfect model scenario so the conclusions can be deceiving if the model is a poor approximation of DGP.

        \paragraph{Stochastic Simulator\\}
        The presence of stochastic element in $f(\cdot)$, and most likely also in $\mathscr{A}$, means that any comparison can only be made in a statistical sense \citep{benekohal1994variability}. The most common approach is to compare the expected value using the usual Monte Carlo estimator, namely the empirical mean of several runs which essentially replaces model theoretical statistic with finite sample estimate \citep{benekohal1991procedure, gagnon2008calibration}. Furthermore, the simulator $f(\cdot)$ can be heteroskedastic so the number of runs required for a given confidence level might vary, such as in the case of under and over-saturated conditions \citep{tian2002variations}. Instead of just point comparison, one might conduct classical hypothesis tests such as the $t$-test \citep{park2003microscopic, toledo2004statistical} for comparing the means of two groups or the Kolmogorov-Smirnov test for comparing two distributions \citep{kim2005calibration}. However such tests might not be applicable if the data is not i.i.d. such as time series data \citep{hourdakis2003practical}. One interesting direction is the use of \emph{variance reduction techniques} \citep{rathi1992variance} to reduce the number of runs required for a given confidence level by exploiting the pseudo random number generator to construct suitable control variates. Aside from multiple evaluations needed, an important difference between a deterministic simulator and a stochastic simulator is the direct accessibility of the likelihood which we will elaborate more in section~\ref{bayesian_computation}.

        \paragraph{Non-Identifiability\\}
        An optimisation approach essentially assumes that model fit is summarised by a single and often scalar loss function. \emph{Model non-identifiability} occurs when many combinations of parameters values result in equally low error \citep{bayarri2004assessing, molina2005statistical} and hence in an error-centric inference, provide equally good fit. It is related to the multi-modality of \eqref{loss_function} and poses a challenge in identifying the supposedly correct optima. Some studies propose a multi-objective approach \cite{karimi2019two} which returns an ensemble of non-dominated solutions lying on the Pareto optimal front and are all optimal in some sense. However choosing a solution among them inevitably requires either domain expertise or some form of scalarisation. 

        We reiterate this as a major difference  between model calibration in engineering and training in ML. Conventional ML model parameters lose interpretability and optimisation is purely driven by monitoring the error without much regards to the estimated parameters values. However for (semi) theory-driven and explanatory models, simply monitoring the error might not be sufficient. If one believes in the credibility of the model, then the \emph{simultaneous convergence of error to an acceptably low threshold and parameters estimates to plausible values} are arguably necessary (but perhaps still not sufficient) conditions for successful model calibration.\\
        
        We end this model calibration section with a reminder on a seemingly harmless but indeed poor practice. Even if the error is low and the parameters estimates converge to plausible values so one is contented with model calibration, further inferences about the reality based on the estimates must be drawn with care. The credibility of such inference crucially depends on the \emph{reliability of the estimates}. \cite{ossen2009reliability} provided an excellent example to illustrate this. For instance, they found that free flow car following parameters cannot be estimated reliably if the trajectory data consists of mostly congested flow conditions. The implication is that during calibration, the model will be insensitive to changes in free flow parameters so a large range of values remain equally plausible and there is hardly any justifiable preferences for point estimators. A point estimate of say free flow speed is highly uncertain and cannot be trusted to be indicative of the actual free flow speed so inferences about drivers' free flow speed should acknowledge such residual uncertainty. This is the motivation of proper uncertainty quantification using Bayesian inference, which will be discussed in section~\ref{uncertainty_quantification_and_propagation}.


\section{Model Validation} \label{model_validation}
This section now address the second step of model validation in model calibration and validation. The commonly adopted definition for model validation as ensuring the model approximates reality sufficiently well unfortunately provides virtually no insights on the procedures and its statistical rationale. In fact, our literature review found significant inconsistencies in the interpretation and consequently implementation of model validation. 

We first distinguish model validation from \emph{model verification} \citep{benekohal1991procedure} which are often mentioned together. Model verification checks for the correctness of computation from its intended purpose, which alone does not guarantee the usefulness of the model. In our subsequent discussions, we will assume model verification has been performed so the model executes as expected and is free from code bugs.

In a typical ML setup, validation is often mentioned in a predictive sense and assessed based on out-of-sample predictive performance. The data is first partitioned into the training and validation set $D = D_{tr} \cup D_{val}, D_{tr} \cap D_{val}=\varnothing$. The model is then trained on $D_{tr}$ and validated by comparing the predictive of the calibrated model with $D_{val}$ which has not been used in fitting the model. However, model validation is problem and context specific \citep{sacks2002statistically} and is certainly not restricted to the aforementioned interpretation ubiquitous in ML. The aim of this section is to distinguish the conceptually different rationales of model validation and to raise awareness on the inconsistencies in the literature.

    \subsection{Scientific vs Statistical Validation}
    We refer to the two fundamentally different philosophies of model validation as \emph{scientific validation} and \emph{statistical validation}. Although they have a common intention of assessing model fit, they differ significantly in their conceptual motivation and hence detailed implementation.
    
    In scientific model validation, model fit is not necessarily measured by out-of-sample predictive performance so the same set of data used for calibration might be subsequently used for validation. A simple example is the analysis of residuals when assessing the model goodness-of-fit. In traffic modelling, a common scientific model validation approach is to compare the model predictions against the observed data for other auxiliary outputs that are not part of \eqref{loss_function}. As alluded to, the full outputs are often of high dimension so model calibration is usually only performed by inspecting a small subset of it. Let $(Y, Y')$ denote the outputs augmented with the auxiliary outputs. Model calibration is first performed by minimising \eqref{loss_function} which involves $Y$ and scientific model validation then assesses model fit by comparing the calibrated model's prediction on $Y'$ with its observed counterpart using possibly the same data\footnote{\cite{toledo2004statistical} adopted a more convoluted approach by fitting meta models for both observed and simulated outputs and conduct statistical tests on the equality of the meta model parameters but it is conceptually the same comparison.}. Some literature whose notion of model validation is in fact scientific model validation includes \cite{benekohal1991procedure, wu2003validation} and \cite{ni2004systematic}. Since $Y$ and $Y'$ are often correlated, information concerning $Y'$ might already be used implicitly when calibrating the model even if it is not directly involved in the loss function. Hence one has to be careful when associating good model predictive performance solely based on scientific model validation. Although such `double dipping' on the same data is forbidden in statistical model validation to be discussed next, it might be sensible if the model is somewhat explanatory. In this case, the simultaneous match on all relevant outputs is perhaps a necessary condition for a validated model. 

    On the other hand, statistical model validation is the more commonly interpretation of model validation. It measures model fit by out-of-sample predictive accuracy. The common interpretation of validation in ML mentioned at the start of section~\ref{model_validation} is more unambiguously referred to as \emph{cross-validation}. Such data partitioning and error comparison might be repeated multiple times and averaged which leads to \emph{k-fold cross-validation}. In fact statistical model validation hinges on the definition of parameters optimality in \eqref{optimal_param_def} and the bias-variance tradeoff mentioned in section~\ref{optimality}. We will illustrate the connection in section~\ref{bias_variance_tradeoff}.
    
    \subsection{Different Notions of Independence}
    Despite the consensus that some notion of independence is required when performing (statistical) validation, the interpretations of such independence vary in the literature and some are arguably flawed. The usual notion of independence is that all data points in $D_{tr}$ and $D_{val}$ are i.i.d.. However, we have observed two other interpretations that might seem sensible at first glace but are not strictly correct from a statistical learning perspective:
    \begin{itemize}
        \item \emph{Independence between outputs}\\
        The first is the (conditional) independence between different output variables. As mentioned before, some researchers such as \cite{park2003microscopic} compare different output variables as part of scientific model validation. This alone does not contradict statistical model validation and assessment of predictive performance if $D_{tr}$ is different to $D_{val}$, and is somewhat an overkill but one might argue it adds even more credibility to the model. The issue arises when the same data is used. On one hand, if $Y$ and $Y'$ are conditionally dependent given $X$, then information for $Y'$ was already used when fitting the model by matching $Y$ so a subsequent good fit to $Y'$ might be hardly surprising. On the other hand, if they are conditionally independent, then there is no reason to expect a model calibrated on $Y$ to be able to predict $Y'$ well.
        \item \emph{Different scenarios}\\
        The second is the difference in the DGP for $D_{tr}$ and $D_{val}$. \cite{park2003microscopic} and \cite{benekohal1991procedure} suggested validating model under `different scenarios'. For example the model is calibrated using morning peak data and validated using evening peak or non-peak data. However, this certainly violates the assumption that the data points are i.i.d.. Intuitively, why should one even expect a model fitted to one scenario to be able to generalise to a completely different scenario? Although this might be sensible for non i.i.d. data as long as the $\theta_i$'s are exchangeable (see section~\ref{parameters}), the different scenarios often imply more prior knowledge so assumption on parameter exchangeability does not hold.
    \end{itemize}

    The discussions above do not attempt to dismiss any existing studies. Our sole intentions are to 1) unify the different interpretations and consequently implementations of model validation and more importantly, 2) elicit the subtle logical flaws and promote statistically rigorous practices. We end our discussion on the interpretations of model validation by acknowledging once more the difference between explanatory and ML models. The former which is the case for most traffic models contains theoretical derivations and hence can potentially \emph{extrapolate}. As such, there is less of a requirement for rigorous statistical model validation before deploying the model. Our hope is that this section served as a cautionary advice for modellers to identify the correct interpretation and procedure for their validation purposes.
    
    \subsection{Bias-Variance Tradeoff} \label{bias_variance_tradeoff}
    This section briefly illustrates how cross-validation is essentially a diagnostic procedure of the bias-variance tradeoff mentioned in section~\ref{optimality}.  Let 
    \begin{align*}
        \epsilon_{tr} &:= \frac{1}{|D_{tr}|}\sum_{(\tilde{x}_i, \tilde{y}_i) \in D_{tr}} \frac{1}{|\Xi_i|}\sum_{\xi_{ij} \in \Xi_i} C_i( M_{\xi_{ij}}(\tilde{x}_{i}, \theta), \: \tilde{y}_{i})\\
        \epsilon_{val} &:= \frac{1}{|D_{val}|}\sum_{(\tilde{x}_i, \tilde{y}_i) \in D_{val}} \frac{1}{|\Xi_i|}\sum_{\xi_{ij} \in \Xi_i} C_i( M_{\xi_{ij}}(\tilde{x}_{i}, \theta), \: \tilde{y}_{i})\\
        \epsilon_{gen} &:= \mathbb{E}_{\mathbf{D}, \xi} [ C(M(\tilde{X}, \theta| \xi), \tilde{Y})]
    \end{align*}
    denote the training error, validation error and generalisation error respectively for a given parameter value $\theta$. In this section, we will implicitly conditioned $\theta$ on the empirical solution to \eqref{loss_function}. Up to random fluctuations, we can expect $\epsilon_{val} \geq \epsilon_{tr}$ so there are two cases:
    \begin{enumerate}
        \item \emph{High bias} occurs when $\epsilon_{val} \approx \epsilon_{tr} >> \epsilon_{th}$. Assuming the reason is not due to an inefficient optimiser getting stuck in local optima which can be diagnosed by high variability of $\epsilon_{tr}$ from multiple optimisation runs, it is an indication of an under-fitting model.  The model and the choice of parameters to be estimated simply do not have the required complexity to express the data. The meaning of bias here is analogous to the aforementioned $\mathbb{E}[\hat{\theta}] - \theta^*$. Even if we could collect so much data that standard error is virtually zero and $\hat{\theta}$ converges to its expectation $\mathbb{E}[\hat{\theta}]$, this is far from the optimal $\theta^*$ so the model will not predict better just by throwing more data at it. A more complex model is needed and one option is to include more arguments to the model as calibration parameters.
        \item \emph{High variance} occurs when $\epsilon_{val} >> \epsilon_{tr} \approx \epsilon_{th}$ which is the more common and feared case of an over-fitting model. This is also more prevalent in the literature because traffic models have become extremely complex and notorious for their large number of parameters\footnote{Worse still, the number of parameters can further scale in a hierarchical fashion such as with different vehicle classes, road types, population of drivers etc.}. Likewise, the meaning of variance here is analogous to the variance of the estimator $\mathbb{E}[(\hat{\theta} - \mathbb{E} [\hat{\theta}])^2]$. Given finite data, any estimate $\hat{\theta}(D)$ might be quite far from its expectation $\mathbb{E}[\hat{\theta}]$ because we have fitted to features that are specific only to $D$. The opposite suggestions to the previous case of high bias hold. The purpose of a regulariser in \eqref{loss_function} is one option to reduce such variance. Furthermore, it becomes clear that the conventional suggestion of `getting more data' is then just an application of the weak law of large number i.e. convergence of $\hat{\theta}$ to $\mathbb{E}[\hat{\theta}]$ in probability as sample size increases.
    \end{enumerate}
    The formal justification of statistical model validation and cross-validation is that $\epsilon_{val}$ is a more unbiased estimator of $\epsilon_{gen}$ than $\epsilon_{tr}$. The out-of-sample predictive performance described by the validation error is a better estimate of the expected model generalisation performance which ultimately defines the optimality as per \eqref{optimal_param_def} that we seek. What one hopes to find is a sweet spot that balances model error and model complexity, supposedly when $\epsilon_{val} \approx \epsilon_{tr} \approx \epsilon_{th}$.

    \subsection{Test Data}
    The solution to \eqref{loss_function} will also depend on various hyperparameters such as $\phi$ that indexes the specific instance of the optimisation algorithm used within its family, and $\lambda$ the regularisation coefficient. Cross-validation can similarly be used to select appropriate values for them. Typical ML pipeline will in fact partition the data into three sets $D = D_{tr} \cup D_{val} \cup D_{te}$ with the last referred to as the test set. This is however rarely adopted or even mentioned in existing model calibration frameworks presumably due to data scarcity. The necessity of $D_{te}$ arises because of the iterative nature of model calibration and validation. The collective choices that the modeller made when refining choices of hyperparameters and obtaining the corresponding parameters estimates can over fit both $D_{tr}$ and $D_{val}$. The model predictive performance on $D_{te}$ which was never used in this iterative process should then be a more unbiased estimate of $\epsilon_{gen}$.


\section{Bayesian Inference} \label{bayesian_inference}
Our discussion on model calibration and validation thus far has focused on an optimisation-based approach centered around the point estimator given by \eqref{loss_function}. In this section, we will discuss Bayesian inference as an alternative framework for parameters estimation. We will outline the relative advantages and disadvantages of a Bayesian framework compared to the simpler optimisation based approach. We will then highlight their connections before summarising the application of Bayesian inference in estimating traffic model parameters.

    \subsection{Bayesian Statistics Background} \label{bayesian_statistics_background}
    We will first provide some background on Bayesian statistics which might be unfamiliar to some readers and refer interested readers to excellent review by \cite{martin2023computing}. Bayesian statistics originated in the mid $17^{\text{th}}$ century which is far earlier than the more widely taught classical statistics nowadays. However it only started gaining popularity in the last few decades because of both advances in computing power and more importantly, development of algorithms to numerically solve the associated computational problems. Bayesian statistics differ from classical (frequentist) statistics foundationally in their definition of probability\footnote{We focus on a Bayesian approach in this paper because it is flexible and directly applies to model calibration and validation. In fact Bayesian and frequentist methods often produce similar results. In the first author's opinion, a Bayesian approach is conceptually simple but computationally challenging while a frequentist approach is often the opposite.}. Bayesians associate probability with subjective belief as opposed to long-run frequency from hypothetical resampling, so uncertainty is associated with \emph{a lack of information}. This subtle change of perspective allows Bayesian statistics to describe unknowns such as parameters as random variables and uncertainty about them using the language of probability.

    The governing equation in Bayesian inference is the Bayes' rule:
    \begin{align}
        p(\theta|D) &= \frac{p(D|\theta) p(\theta)}{p(D)} \label{bayes_rule}\\
                    &\propto p(D|\theta) p(\theta).
    \end{align}
    The \emph{prior} distribution $p(\theta)$ describes the uncertainty about the parameters before observing any data. The \emph{likelihood} $p(D|\theta)$ is the sampling distribution i.e. the
    statistical model that approximates the DGP. Most importantly, the \emph{posterior} distribution $p(\theta |D )$ describes the remaining uncertainty about the parameters after observing the data. If the data provides information about the values of $\theta$, the intuition is that observing $D$ reduces our uncertainty about $\theta$ so $p(\theta|D)$ concentrates onto a small region of $\Theta$ compared to $p(\theta)$.

    \subsection{The Advantages: Flexibility and Statistical Coherence} \label{flexibility_and_statistical_coherence}
    A Bayesian framework provides a simple yet flexible and statistically coherent methodology to combine different sources of uncertainties when estimating unknowns.

    \subsubsection{Uncertainty Quantification and Propagation} \label{uncertainty_quantification_and_propagation}
        The estimator in optimisation based approach, namely the solution to \eqref{loss_function} is ultimately just a point estimator\footnote{Even if the optimisation algorithm is an ensemble method that returns a number of plausible parameters values when the algorithm terminates, one has to be careful about its probabilistic interpretation which in this case only describes the Monte Carlo error of the optimisation algorithm.}. When a model is calibrated and validated using an optimisation-based approach, the optimal point estimate $\hat{\theta}(D)$ is fixed for all subsequent predictions. This is potentially problematic for two reasons. The first concerns the reliability of the estimate when drawing subsequent inference as mentioned at the end of section~\ref{practical_considerations} and the second concerns the \emph{predictive uncertainty}. Let $D' = (X', Y') \notin \mathbf{D}$ and $d' = (x', y')$ denote a new data point and its realisation respectively, then the prediction for $Y'$ is given by $M(x', \hat{\theta}(D))$. This ignores the uncertainty in $\hat{\theta}(D)$ itself which means the predictive uncertainty for $Y'$ will be under-estimated. A Bayesian approach addresses both of these issues. 
        
        The posterior $p(\theta|D)$ is the answer to the first concern of \emph{uncertainty quantification} by expressing the remaining uncertainty about $\theta$ with probability distributions. The posterior returns a distribution from which one can compute answers to almost any question one can ask about $\theta$. Some examples are:
        \begin{itemize}
            \item Point estimator such as the \emph{posterior mean}
            \[
            \mathbb{E}[\theta|D] = \int_{\Theta}\theta p(\theta|D) d\theta
            \]
            \item Region/interval estimator such as a $100(1-\alpha)\%$ credible region\footnote{In one dimension, this is also referred to as credible interval. Credible interval is conceptually different to confidence interval because of the different interpretation of probabilities but they can coincide.} i.e. $\Theta' \subseteq \Theta$ s.t. 
            \[
            \int_{\Theta'}p(\theta|D) d\theta = \int_{\Theta} \mathbf{1}_{\Theta'}(\theta)p(\theta|D) d\theta = 1-\alpha
            \]
            where $\mathbf{1}_{\Theta'}(\theta) := 1$ if $\theta \in \Theta'$ and 0 otherwise is the indicator function.
            \item Probabilities such as that for $\theta$ belonging to a certain region/interval $\Theta''\subseteq \Theta$
            \[
            p(\theta \in \Theta'') = \int_{\Theta} \mathbf{1}_{\Theta''}(\theta) p(\theta|D) d\theta.
            \]
        \end{itemize}
        Note that Bayesian inference ideally uses the whole posterior distribution to conduct inference, hence any point/interval estimator is merely a convenient summary to convey the shape of the posterior.

        The second concern of \emph{uncertainty propagation} is answered by the \emph{posterior predictive distribution}, which in our notation is given by 
        \[
        p(Y'| D, x') = \int_{\Theta} M(x', \theta) p(\theta | D) d\theta. \label{posterior_predictive_eq}
        \]
        The Bayesian approach to incorporate uncertainty due to \emph{nuisance parameters}\footnote{Nuisance parameters are those that are not of interest. In prediction, the focus is on $Y'$ so $\theta$ is hence a nuisance parameter because it was just a medium to get to $Y'$.} is to marginalise them out and this is precisely what the posterior predictive distribution performs. One interpretation is that a Bayesian approach uses all possible parameters values weighted by their posterior densities which describe the relative plausibility of each parameter value. The posterior predictive distribution hence properly propagates the uncertainty in $\theta$ to the uncertainty in $Y'$.

        \subsubsection{Incorporation of Domain Knowledge}
        Although the posterior is often the main quantity of interest in Bayesian inference, the prior $p(\theta)$ also has its role. It provides a natural way to incorporate domain expertise to directly influence the posterior.  A common challenge when estimating physically meaningful parameters, is the tradeoff between the values supported by domain expertise which produce worse fit and the better-fitting but perhaps less plausible values. When choosing a prior, the modeller chooses how much probability mass to be placed on each open set around each point in $\Theta$ before seeing the data. This means a modeller can assign more prior mass to values they are more inclined believe in for whatever reason. This could be based on the physical meaning of the parameters, the default values, estimates derived from observing auxiliary variables, or even results based on previous studies such as the meta-analysis conducted in \cite{xu2015calibration}. The prior will be more sharply peaked around these values and it will require more data to overthrow such prior belief.

        Although the prior is often criticised as subjective choices and should be kept uninformative to ``let the data speak for itself" via the likelihood, the likelihood also inevitably involves assumptions on the statistical model. An informative prior in fact plays the role of a regulariser in \eqref{loss_function} (see section~\ref{section:connection_to_optimisation} and can significant improve the well-posedness of the inference task.

        \subsubsection{Hierarchical Modelling}
        We alluded to hierarchical modelling in several sections where the parameters to be estimated has a natural hierarchical structure which is common in traffic modelling. The joint estimation approach adopted by \cite{hoogendoorn2010generic} is in fact one such cases. Another more obvious example is \cite{huang2010multilevel} who used hierarchical generalised linear model for crash prediction. Suppose the car following model example is applied to a platoon of many vehicles and resulted in many leader/follower pairs. It is often more meaningful not to estimate say the reaction time of a particular follower but instead the statistical parameters that describe the driver reaction time distribution such as the mean. The parameters exchangeability assumption mentioned in section~\ref{parameters} is once again the rigorous reason behind hierarchical modelling. 
        
        Instead of estimating a common $\theta$ for each of the $n$ data point which corresponds to the trajectory data for each pair of leader and follower, we explicitly estimate a $\theta_i$ for each of them and more importantly the common $\psi$ that governs their distributions. The model is extended to
        \begin{align*}
            \psi &\sim p(\psi)\\
            \theta_i|\psi &\overset{\mathrm{iid}}{\sim}p(\theta_i|\psi)\\
            \tilde{Y}_i|\theta_i, \Tilde{X}_i &\mathrel{\dot\sim} M(\tilde{Y}_i|\tilde{X}_i, \theta_i)
        \end{align*}
        for $i= 1, 2, ..., n$. For brevity let $\theta = \{\theta_i\}_{i=1}^{n}$ denote all the individual parameters.
        
        The optimisation based approach is not well-suited for estimating $\psi$ because $\psi$ is related to the data only through the unobserved $\theta$. The simple plug-in approach where the $\theta_i$'s are first estimated by solving \eqref{loss_function} and these point estimates $\hat{\theta}_i$'s are used to perform a subsequent estimation of $\psi$, will similarly underestimate the uncertainty in $\psi$, see section~\ref{uncertainty_quantification_and_propagation}.
        
        A Bayesian approach naturally handles the estimation for such hierarchical model by viewing $p(\psi)$ as \emph{hyperprior} and applying Baye's rule
        \[
        p(\psi, \theta |D) \propto p(D | \theta) p(\theta|\psi) p(\psi)
        \]
        as per usual. We obtain the joint posterior $p(\psi, \theta |D)$ of the individual parameter $\theta_i$ and the population parameter $\psi$. Since the parameter of interest is $\psi$ and $\theta$ is hence just a nuisance parameter, a Bayesian approach can similarly integrate them out to obtain the \emph{marginal posterior}
        \[
        p(\psi |D ) = \int p(\psi, \theta| D) d\theta
        \]
        which is an honest representation of the uncertainty in estimating the higher level parameter $\psi$.

        \subsection{The Disadvantages: Computations} \label{bayesian_computation}
        Bayesian inference provides much more informative estimation results at the expense of computations. Given a choice of prior $p(\theta)$ and a computable likelihood $p(D|\theta)$, the posterior $p(\theta|D)$ is only known up to a normalising constant because the denominator of the RHS in \eqref{bayes_rule} namely $p(D)$ is often expensive or impossible to compute. 
        
        Furthermore, most quantities of interest such as posterior mean, credible intervals, posterior probabilities, posterior predictive distribution, marginal posterior etc (we have put them in a similar form to emphasise this) in Bayesian inference requires one to compute expectations. These are often analytically intractable integrals of the form
        \begin{align}
            J &=  \mathbb{E}_{\pi(\theta)}[g(\theta)] \label{eq:expectation}\\
              &= \int_{\theta \in \Theta} g(\theta) \pi(\theta) d\theta
        \end{align}
        where $h(\theta)$ is some function of $\theta$ and $\pi(\theta)$ is a target distribution which is typically the posterior in Bayesian inference. While quadrature rules can be used to numerically integrate \eqref{eq:expectation}, they quickly become infeasible because of exponential complexity in $p$, the dimension of $\theta$.
         
        Instead, Bayesian computations usually involved constructing a Monte Carlo estimator to \eqref{eq:expectation}, namely $\hat{J}_n = \frac{1}{n} \sum_{i=1}^n h(\theta_i)$ where $\theta_i \sim \pi(\theta)$. The estimator $\hat{J}$ converges to $J$ in probability as $n \rightarrow \infty$ by the weak law of large number and such convergence does not explicitly depend on the dimension. It turns a difficult problem of integration into one of sampling from a distribution which unfortunately is still a difficult task. However, development of popular and universally applicable sampling algorithms such as \emph{Markov chain Monte Carlo} (MCMC) has provided tools to perform such computations. MCMC often only requires one to able evaluate the unnormalised $\pi(\theta)$ pointwise which is a rather non-restrictive assumption that suits Bayesian inference well. 
        
        The general idea of MCMC algorithms is to implicitly construct a Markov chain with $\pi(\theta)$ as its limiting distribution. Recall a discrete time \emph{Markov chain} is a sequence of random variables $Z_1, Z_2, ...$ with the Markov property
        \[
        p(Z_{j+1} = z | Z_1 = z_1, Z_2 = z_2, ... , Z_j = z_j) = p(Z_{j+1} = z | Z_j = z_j),
        \]
        i.e. the transition probabilities between states only depend on the previous state instead of the entire history.
        Let $\Theta$ denote the state space of the Markov chain which is the original parameter space and let $q(\theta, A)$ denote the \emph{transition kernel} of the Markov chain from $\theta$ to any measurable subset $A \subseteq \Theta$. Then the Markov chain is ergodic to $\pi(\cdot)$ if 
        \[
        \lim_{n \rightarrow \infty}||q^n(\theta, A) - \pi(A)||_{TV} = 0 \quad \forall \theta \: \text{and} \: A,
        \]
        and $||.||_{TV}$ is the total variation distance\footnote{The total variation distance between two distributions $\pi_1$ and $\pi_2$ is $||\pi_1 - \pi_2||_{TV} = \sup_A |\pi_1(A) - \pi_2(A)|$.}. In other words, the states obtained by simply iterating the Markov chain forward will eventually be the same as samples from $\pi(\cdot)$. MCMC is similar to an optimisation algorithm by iteratively moving closer to its goal, be it sampling from a given distribution or locating a local optimum, because it is difficult to do so directly. By construction the samples will be correlated but the Markov chain analog of the law of large number and central limit theorem still provide theoretical guarantees on their validity for $\hat{J}_n$. Once the samples are obtained, $\hat{J}_n$ can easily be evaluated as the empirical mean of the samples.

        The literature for state-of-the-art MCMC algorithms is vast and a fast evolving field. We refer readers to specialised statistical texts such as \cite{givens2012computational, gelman2013bayesian} and \cite{craiu2014bayesian} for a more thorough review on MCMC. The main drawback is that MCMC is often  computationally intensive and requires a large number of model evaluations (often of order $10^3 - 10^6$ or more) for the chain to converge and fully explore the posterior. More efficient state-of-the-art samplers such Hamiltonian Monte Carlo and in particular the No-U-Turn sampler \citep{hoffman2014no} requires gradients which are often not accessible for the traffic models and hence cannot be readily applied. Furthermore, the likelihood $p(\theta|D)$ is not even directly computable if $f(\cdot)$ is a stochastic simulator. In this case, $f(\cdot)$ is actually a sampler for the underlying distribution induced by the stochastic model but we cannot compute the density directly. One is then forced to either estimate the density with many draws from the simulator, or resort to likelihood free methods such as approximation Bayesian computation \citep{beaumont2019approximate, martin2023approximating}, both of which can quickly become even more computationally expensive. 

    \subsection{Connection to Optimisation Based Approach} \label{section:connection_to_optimisation}
    Although optimisation based approach and Bayesian inference might appear as different approaches to parameters estimation, the two are in fact conceptually similar. The difference is that the focus of an optimisation approach is on \emph{prediction} while that of a Bayesian approach is on \emph{inference}. This ultimately results in their different methodologies and treatment of uncertainty. Unlike optimisation, the primary operation in Bayesian inference is to compute expectation and hence integration. 
    
    If a Bayesian must provide a point estimator, its optimisation analog is the \emph{maximum a posterior} (MAP) estimator 
    \begin{align*}
    \hat{\theta}_{MAP} &= \argmax_{\theta \in \Theta} p(\theta | D)\\
                    &= \argmax_{\theta \in \Theta} \log p(\theta | D)\\
                    &= \argmax_{\theta \in \Theta} \log p(D|\theta) + \log p(\theta)
    \end{align*}        
    which is similar to \eqref{loss_function} both in their forms and conceptual roles. The log-likelihood $\log p(D|\theta)$ has the role of the cost function and measures the modelling error i.e. how well the model output matches the observed output for a given parameter value. The log-prior $\log p(\theta)$ has the role of the regulariser and measures model complexity. A concentrated prior with small variance has the same effect as a high $\lambda$ and hence model complexity is penalised more. Indeed, many models such as ridge regression \citep{lindley1972bayes} can be shown to have a Bayesian interpretation. 
    
    One interpretation of a loss function in \eqref{loss_function} is that it can be viewed a generalisation of methods for constructing a point estimator. A somewhat arbitrarily chosen cost function replaces the likelihood and similarly a regulariser replaces the prior. Although the optimisation based approach is more flexible for constructing point estimator in this sense, the components do not necessarily have probabilistic interpretations and hence statistical coherence is lost. As a result, optimisation-based practitioners are concerned with their choices of estimator (see section~\ref{numerical_optimisation}) while a Bayesian approach only ever has one estimator, the whole posterior distribution. It is not that Bayesian inference is free from choices but rather it is explicit and honest about its assumptions when eliciting the prior and the likelihood, from which everything else must be logical consequences of. 

    \subsection{Application in Traffic Modelling} \label{application_bayesian_inference}
    A full Bayesian approach is relatively uncommon in traffic modelling \citep{hazelton2010bayesian}. Aside from possibly requiring stronger statistical background and familiarity with probabilistic programming, the most obvious reason is its computationally expensive nature as discussed in section~\ref{bayesian_computation}. A number of studies \citep{tebaldi1998bayesian, li2005bayesian, hazelton2010bayesian, perrakis2015bayesian, parry2013bayesian} applied Bayesian inference for OD estimation, often using a Poisson mixture model with a log-link function i.e. the log of the Poisson rate parameter is a linear combination of the predictors. \cite{hazelton2010bayesian} replaced the link function with a logit route choice model but the structure of the Poisson mixture model remains largely unchanged. Since the posterior is not analytically tractable, MCMC is often necessary for exact inference but some approximations were also attempted to speed up computations. For instance, \cite{li2005bayesian} used the expectation maximisation algorithm to obtain a MAP estimate and performed rescaling to obtain posterior variance while \cite{perrakis2015bayesian} used integrated nested Laplace approximation which approximates the posterior distribution with a Gaussian centered at the mode of the posterior and has covariance equal to the inverse Hessian evaluated at the mode. However, the Poisson mixture model belongs to the popular generalised linear model with closed form representation so inference does not strictly involve evaluating a complex black box simulator.
    
    Application of Bayesian inference to such complex simulator without closed form has attracted relatively little attention. \cite{van2015general} attempted to compute the evidence which is the denominator on the RHS in \eqref{bayes_rule} conditioned on different car following models, as a method for comparing them\footnote{Model selection and model averaging are further examples to illustrate the flexibility of Bayesian inference, but we did not have space to discuss them in more detail.}. Computing evidence involves difficult integrals so they assumed normality and used Laplace approximation. \cite{abodo2019strengthening} constructed a hierarchical model and compared different pooling approaches when estimating the individual behavioural parameters in car following model. \cite{ji2021bayesian} attempted to infer the delay time and (air/rolling) drag resistance parameters when a connected autonomous vehicle is following a connected human vehicle. 
    
    \cite{bayarri2004assessing} and \cite{molina2005statistical} appear to be the only studies that applied a Bayesian approach for a stochastic network traffic simulation model, which in their case was CORSIM. Their goal was to infer demand (number of counts entering the network from external sources) and turning probabilities (the probabilities of making left/through/right turn at an intersection) based on observed count data. Their successful application of Bayesian inference hinges on the crucial fact that the parameters only enter CORSIM in a relatively simple structure. In particular, CORSIM models vehicle turns independently of its previous history and other vehicles which induces conditional independence. Hence the simulator can be well approximated by a stochastic network that models counts and turning probabilities with Poisson and multinomial distributions respectively. As a result, MCMC can run much more efficiently without the need to call the expensive simulator at every iteration.


\section{Gaps for Future Research} \label{improvements}
We end by listing some gaps and future research directions that we believe will be beneficial to advancing the state-of-the-art practices for model calibration and validation. Firstly, the choices involved in constructing a point estimator for \eqref{optimal_param_def}, namely the outputs, the cost function and the optimisation algorithm, are most certainly context and problem specific. Benchmarking and comparison studies such as those performed by \cite{ossen2008validity, kesting2008calibrating, punzo2012can, ciuffo2013no} and \cite{punzo2021calibration} will provide useful guidelines on empirically well performing combinations on similar problem. An example of such problem specific parametrisation the recommendation by \cite{punzo2021calibration} to use relative spacing as outputs instead of relative velocity when calibrating car following models.

In addition, our discussion reveals that the motivation of model calibration and validation is largely based on statistical inference. A wider adoption of proper statistical inference methodologies such as Bayesian inference, and development of feasible computational methods possibly with approximations, will help to promote statistically rigorous and correct procedures. Most notably, modellers can quantify the reliability of their parameters estimates, which is necessary if one is to make further inference based on them. We stress that this does not dismiss the value of an optimisation based point estimator and also Bayesian inference is not always the ideal choice. However, the appropriate parameters estimation methodology is such an important consideration yet is not discussed much in the existing literature filled with applications of optimisation based approaches.

Lastly, our discussions thus far have largely ignored the fact that $M(\cdot)$ is only an approximation to the DGP, which we referred to as model imperfection error. Such error is not easily separated from observation error because only the their sum is observed. Some effort has been made to account for such modelling bias, the most notable of which is the seminal paper by \cite{kennedy2001bayesian} who used Gaussian process to model the bias due to model imperfection error. We refer interested readers to dedicated text such as \cite{kennedy2001bayesian} and \cite{bayarri2007framework} for a more thorough discussion on accounting for model bias.


\section{Conclusions} \label{conclusions}
In this paper, significant inconsistencies and impactful malpractices for model calibration and validation were identified when reviewing existing transport modelling literature. The terminologies rampant in the existing literature are often vague so the language from statistical inference was necessary for our general and statistically consistent formulation of model calibration and validation.

Our main findings are summarised below. Firstly, the modelling aim is what distinguishes between inputs and parameters while the model credibility and parameter observability are what distinguish between parameters to be estimated versus treated as known. Secondly, the parameter exchangeability is often the actual assumption behind model calibration and validation as opposed to say i.i.d. data. Thirdly, the plethora of model calibration procedures and frameworks in the existing literature can be unified in terms of estimators from statistical inference. Fourthly, although model calibration and validation also seek best fitting parameter value that minimises the generalised error as per bias-variance tradeoff, they differ from the conceptually similar training and cross validation in machine learning. The model of interest is often explanatory so
1) the simultaneous convergence of calibration error to an acceptable threshold and parameters estimates to plausible values is arguably a necessary condition, 2) it can be much more rigid so an achievable error threshold is not always known a priori which makes setting an acceptable error threshold challenging and 3) model validation is not necessarily restricted to statistical validation. Lastly, although the common optimisation based point estimator is a sensible approach to model calibration, statistical inference methodologies such as Bayesian inference should be considered if the purpose of the model is not just prediction but also inference.

It is our hope that this research will help the transport modelling community to better understand the statistical rationale of model calibration and validation, and see how the existing practices fit together. Most importantly, we hope that this paper will help practitioners avoid seemingly sensible but in fact flawed practices.


\section*{CRediT authorship contribution statement}
\textbf{ST}: Conceptualization, Formal analysis, Investigation, Writing - original draft, Writing - review \& editing, Visualization;
\textbf{TL}: Writing - review \& editing, Visualization;
\textbf{TS}: Supervision, Writing - review \& editing;
\textbf{YS}: Supervision, Writing - review \& editing, Funding acquisition;
\textbf{MS}: Supervision, Writing - review \& editing.

\section*{Acknowledgement}
ST is supported by an Australian Government Research Training Program Scholarship at the University of Western Australia, a PATREC Intelligent Transport Systems Postgraduate Research Scholarship, and an iMOVE CRC Postgraduate Research Scholarship funded by iMOVE CRC and supported by the Cooperative Research Centres program, an Australian Government initiative.  MS
acknowledges the support of the Australian Research Council through the Centre for Transforming Maintenance through Data Science (grant number IC180100030),
funded by the Australian Government. MS is also supported by the ARC Discovery Grant DP200102961, funded by the Australian Government.


\clearpage
\renewcommand{\thesection}{\alph{section}}
\section*{Appendix A} \label{appendix_a}

\begin{table}[htbp!]\caption{Summary of Notations}
\small
\begin{tabular}{l p{0.9\textwidth}}
\toprule
$\mathscr{A}_{\phi}(\zeta)$ &family of optimisation algorithm indexed by hyperparameter $\phi$ and a stochastic element $\zeta$ \\
$a_n(t)$ &acceleration of the $n^{\text{th}}$ vehicle in CF model \\
$C(\cdot,\cdot)$ &cost function which is a distance measure between simulated output and observed output\\
$\mathbf{D}$ &= $\{(\tilde{X}_i, \tilde{Y}_i)\}$, random vector of data\\
$D$ &= $\{(\tilde{x}_i, \tilde{y}_i)\}$, observed data and a particular realisation of $\mathbf{D}$\\
$D_{tr}, \: D_{val}, \: D_{te}$ &training, validation and test data respectively\\
$D_i$ &= $(\tilde{X}_i, \tilde{Y}_i)$, the tuple of random vector for the $i^{\text{th}}$ data point\\
$d_i$ &= $(\tilde{x}_i, \tilde{y}_i)$, particular realisation of the $i^{\text{th}}$ data point\\
$f_{\mu}(X, \xi)$ &dynamic model with dynamic parameters $\mu$ and a stochastic element $\xi$\\
$h(\cdot|X, Y, \sigma^2)$ &the stochastic process of data collection\\
$M(\cdot|X, \theta)$ &the model which consists of the dynamic model and the noise model\\
$p(\cdot)$ &a generic notation for probability distribution / probability density / statistical model\\
$R$ &reality which refers to the joint distribution $p(X,Y)$ or the conditional distribution $p(Y|X)$\\
$R(\theta)$ &regulariser in \eqref{loss_function}\\
$s_n(t)$ &displacement of the $n^{\text{th}}$ vehicle in CF model \\
$T(\mathbf{D})$ &a statistic which is a function of the data\\
$t_c, t_f$ &the critical gap and followup headway parameters in roundabout capacity model\\
$v_n(t)$ &velocity of the $n^{\text{th}}$ vehicle in CF model \\
$X, \: x$ &random vector of inputs and its particular realisation\\
$\tilde{X}, \: \tilde{x}$ &random vector of observed inputs and its particular realisation\\
$\hat{X}$ &random vector of estimated inputs, an estimator of $X$\\
$Y, \: y$ &random vector of outputs and its particular realisation\\
$\tilde{Y}, \: \tilde{y}$ &random vector of observed outputs and its particular realisation\\
$\hat{Y}$ &random vector of estimated outputs, an estimator of $Y$\\
$Z, \: z$ &denotes a generic random variable and its particular realisation\\
$\epsilon_{gen}$ &generalisation error\\
$\epsilon_{tr}, \: \epsilon_{val}, \: \epsilon_{te}$ &training, validation and test error respectively\\
$\zeta$ &stochastic element in the optimisation algorithm $\mathscr{A}$\\
$\Theta$ &feasible parameter space\\
$\theta$ &$= (\mu, \sigma^2)$, parameters to be estimated\\
$\theta_i$ &individual parameter for modelling $D_i$\\
$\theta^*$ &optimal parameter value defined in \eqref{optimal_param_def} which minimises the generalisation error\\
$\theta^c$ &auxiliary parameters that conditioned on specific values and not estimated\\
$\hat{\theta}$ &a generic notation for an estimator or estimate of $\theta$\\
$\hat{\theta}(\mathbf{D})$ &an estimator of $\theta$\\
$\hat{\theta}(D)$ &an estimate of $\theta$ and a particular realisation of $\hat{\theta}(\mathbf{D})$\\
$\lambda$ &regularisation coefficient in \eqref{loss_function}\\
$\mu$ &dynamic parameters in the dynamic model $f(\cdot)$, often the subject of calibration\\
$\xi$ &stochastic element in the dynamic model $f(\cdot)$\\
$\sigma^2$ &noise parameters\\
$\phi$ &hyperparameter that indexes a specific instance of an optimisation algorithm\\
$\psi$ &hyperparameter that parametrise the distribution of $\theta_i$\\

\bottomrule
\end{tabular}
\label{tab:TableOfNotationForMyResearch}
\end{table}
\noindent


\clearpage

\bibliographystyle{agsm}
\bibliography{references.bib}

\end{document}